\documentclass[12pt]{article}

\usepackage{graphicx}
\usepackage{amsmath}
\usepackage{amsfonts,amssymb}
\usepackage{latexsym,multibox,amssymb}
\begin{document}
\def\ygp(#1){{\tiny\Yvcentermath1 $\yng(#1)$}}
\newcommand{\mf}{\mathfrak}
\newcommand{\oh}{O(h^\infty)}
\newcommand{\dx}{\partial_x}
\newcommand{\ft}{\footnotesize}
\newcommand{\dy}{\partial_y}
\newcommand{\dt}{\partial_t}
\newcommand{\dz}{\partial_z}
\newcommand{\dxi}{\partial_\xi}
\newcommand{\deriv}[2]{\frac{\partial #1}{\partial #2}}
\newcommand{\ddt}{\frac{d}{dt}}
\newcommand{\lie}{{\cal L}}
\newcommand{\hgt}{\textrm{ht}\,}
\newcommand{\pscal}[2]{\langle #1,#2\rangle}
\newcommand{\ham}[1]{\mathcal{X}_{#1}}
\newcommand{\Lie}[1]{\mathfrak{#1}}
\newcommand{\fourier}{\mathcal{F}_h}
\newcommand{\fouriero}{\mathcal{F}}
\newcommand{\re}{\mathfrak{R}}
\newcommand{\im}{\mathfrak{I}}
\def\pa#1{\left(#1\right)}
\newcommand{\phy}{\varphi}
\newcommand{\epsi}{\varepsilon}
\newcommand{\bep}{\mbox{\boldmath{$\epsilon$}}}
\newcommand{\bc}{\mathbf{c}}
\newcommand{\om}{\omega}
\newcommand{\al}{\alpha}
\newcommand{\la}{\lambda}
\newcommand{\be}{\beta}
\newcommand{\ga}{\gamma}
\newcommand{\de}{\delta}
\newcommand{\restr}{\upharpoonright}
\newcommand{\trsp}{\raisebox{.6ex}{${\scriptstyle t}$}}
\newcommand{\limi}[1]{\displaystyle \lim_{#1}}
\newcommand{\tr}{\textrm{tr}\;}
\newcommand{\demo}[1][$\!\!$]{\noindent\textbf{Proof }\textsl{#1}. }
\newcommand{\gener}[1]{\langle #1 \rangle}
\newcommand{\ind}{\text{\it {\em Ind }}}
\newcommand{\coker}{\text{\it {\em Coker }}}
\newcommand{\ch}{\textrm{ch}\,}
\newcommand{\sch}{\textrm{sch}\,}
\newcommand{\ts}{\mspace{1.2mu}}
\newcommand{\ff}{\emph{focus-focus}}
\newcommand{\U}{\mathcal{U}}
\newcommand{\M}{\mathcal{M}}
\renewcommand{\L}{\mathcal{L}}
\newcommand{\bmu}{\mbox{\boldmath{$\mu$}}}
\def\Lie(#1){\Li <#1>}
\def\SLie(#1){\SLi <#1>}

\renewcommand{\theequation}{\thesection.\arabic{equation}}

\thispagestyle{empty}
\setcounter{page}{0}

{\hfill{\tt hep-th/1007.5241}}
{\hfill{LPT-ENS/09.17}}

\vspace{.5cm}

\begin{center} {\bf \large  $E_{11},$ Borcherds algebras and maximal supergravity }

\vspace{.5cm}

Marc Henneaux$^{1,2,3}$, Bernard L. Julia$^{1}$ and J\'er\^ome Levie$^{1}$

\footnotesize
\vspace{.5 cm}

${}^1${\em Laboratoire de Physique th\'eorique de l'Ecole Normale Sup\'erieure\footnote{Unit\'e mixte de recherche (UMR 8549) du CNRS et de l'ENS,
associ\'ee \`a  l'Universit\'e Pierre et Marie Curie et aux F\'ed\'erations de recherche
FR684 et FR2687.}, 24 rue Lhomond, 75231 Paris CEDEX, France}

\vspace{.2cm}

${}^2${\em Universit\'e Libre de Bruxelles and International Solvay Institutes, ULB-Campus Plaine CP231, 1050 Brussels, Belgium}
\vspace{.2cm}

${}^3${\em Centro de Estudios Cient\'{\i}ficos (CECS), Casilla 1469, Valdivia, Chile}\\

\vspace {0.5cm}

{\tt
henneaux-AT-ulb.ac.be, bjulia-AT-lpt.ens.fr, jlevie-AT-lpt.ens.fr
}

\end{center}

\vspace {1cm}

\centerline{\bf Abstract}
\noindent
The dynamical $p$-forms of torus reductions of maximal supergravity theory have been shown some time ago to possess remarkable algebraic structures. The set (``dynamical spectrum") of propagating $p$-forms has been described as a (truncation of a) real Borcherds superalgebra $\mf{V}_D$ that is characterized concisely by a Cartan matrix which has been constructed explicitly for each spacetime dimension $11 \geq D \geq 3.$ In the equations of motion, each differential form of degree $p$ is the coefficient of a (super-) group generator, which is itself of degree $p$ for a specific gradation (the $\mf{V}$-gradation). A slightly milder truncation of the Borcherds superalgebra enables one to predict also the ``spectrum'' of the non-dynamical $(D\! - \! 1)$ and $D$-forms. The maximal supergravity $p$-form spectra were reanalyzed more recently by truncation of the field spectrum of $E_{11}$ to the $p$-forms that are relevant after reduction from $11$ to $D$ dimensions.  We show 
in this paper how the Borcherds description can be systematically derived from the split ("maximally non compact") real form of $E_{11}$ for $D \geq 1.$ This explains not only why both structures lead to the same propagating $p$-forms and their duals for $p\leq (D\! -\! 2),$ but also why one obtains the same $(D\! -\! 1)$-forms and "top" $D$-forms. The Borcherds symmetries  $\mf{V}_2$ and $\mf{V}_1$ are new 
too. We also introduce and use the concept of a presentation of a Lie algebra that is covariant under a given subalgebra.

\newpage

\section{Introduction}
\setcounter{equation}{0}

\subsection{BKM/Borcherds presentation of $p$-form superalgebras}

$p$-form gauge fields are well known to play a central role in supergravity. Dynamical (alias propagating) $p$-forms ($p \leq D\! -\! 2,$ where $D$ is the spacetime dimension) are necessary for the matching of bosonic and fermionic degrees of freedom. But non-dynamical $p$-forms have also an interesting physical meaning. Indeed, $(D\! -\! 1)$-forms are related to gaugings and cosmological deformations, while (some) $D$-forms couple to space filling branes.

Building on the earlier work of \cite{Cremmer:1997ct,Cremmer:1998px}, it was shown in \cite{HenryLabordere:2002dk,HenryLabordere:2003rd} in spacetime dimension $D \geq 3$ that the spectrum of the dynamical $p$-forms of maximal supergravity as well as  their multiplicities are in one-to-one correspondence with the Cartan generators (but one) and the positive roots of $\mf{V}$-degree at most $(D\! -\! 2)$ of a Borcherds/BKM (super-)algebra\footnote{BKM superalgebras are defined in the Appendix {\bf \ref{DefinitionBorcherds}}. Following T.~Gannon we may use the initials BKM of V.~Kac, R.~Moody as well as R.~Borcherds to denote these superalgebras.}. This $\mf{V}$-degree corresponds to  a particular gradation of the BKM-superalgebra called $\mf{V}$-gradation. In the  supersigma model rewriting of the equations as a self-duality condition, $p$-form potentials are coupled to Borcherds generators of $\mf{V}$-degree $p.$ $\mf{V}$ -- also written V -- stands for vertical (by opposition to the horizontal Cartan degrees of U-dualities). Furthermore,  V-degree truncations of a parabolic subalgebra of the Borcherds superalgebra that contains $E_{11-D}$ act as symmetries of the supergravity field equations (here $E_{11-D}$ is the $U$-duality group ($\mf{U}_D$) in $D$ dimensions and we shall review parabolic subalgebras in the next section). Because these symmetry superalgebras contain the $U$-duality transformations as subalgebras of (form- and V-) degree preserving transformations, they were named $V$-duality in \cite{Julia:2005wg}. This approach is bottom up and one obtains  symmetries of equations or actions. The chiral model with $0$-forms (scalar fields) taking values in a compact or non compact Lie group generalizes to maps into a (super-) group encoding $p$-forms ($p\ge 0$). 

Although we shall consider here only bosonic fields (in the usual sense), superalgebras do appear when odd degree forms are coupled with odd V-degree Borcherds generators, they will be called fermionic below. 

The most spectacular case is perhaps that of type $II_B.$  The Borcherds superalgebra $\mf{V}_{10_B}$ that 
controls the form spectrum of type $II_B$ supergravity turns out to be a rank 2 Borcherds algebra (without fermionic root) defined as explained in the  Appendix  by  the following Cartan matrix \cite{HenryLabordere:2002dk}
\begin{equation}A_{II_B} = \left(
     \begin{array}{cc}
       0 & -1 \\
       -1 & 2 \\
     \end{array} 
   \right) \label{IIB} .
\end{equation} 
As observed in \cite{HenryLabordere:2002dk}, this Borcherds algebra was considered previously 
in \cite{Slansky:1992dx} with different goals in mind and so may be called the Slansky algebra.  In the supergravity context, the null simple root $\alpha_0$ is associated to a "$2$-form generator" 
$e_{\alpha_0},$ while the real root $\alpha_1$ is associated to a scalar generator $e_{\alpha_1}.$ 
By this we exemplify that on the (super-) Borcherds side the fields multiplying the generators are differential forms on spacetime of degree equal to the V-degree attributed  to the generators; this is a stronger restriction than the usual supergroup rule that uses only a $\mathbb{Z} /2\mathbb{Z}$ gradation which is empty in a purely bosonic situation. 
Together with the corresponding Cartan generator $h_{\alpha_1}$ and the lowering operator $f_{\alpha_1},$ the raising operator $e_{\alpha_1}$ generate the type $II_B$ $\mf{sl}\pa{2,\mathbb{R}}$ U-duality symmetry.

The positive roots of the Slansky algebra were recursively constructed in \cite{Slansky:1992dx} using the denominator formula for Borcherds algebras, beyond the height sufficient for our purpose, which was then the study of propagating form-fields.  The $\mf{sl}\pa{2,\mathbb{R}}$-transformation properties of the associated root vectors were also given there (actually $\mf{su}\pa{2}$ representations, as the author of \cite{Slansky:1992dx} considered the compact version).  We reproduce this information in Table {\bf \ref{slansky1}} up to form degree 8 (the roots up to that level are all non-degenerate).
\begin{table}
\caption{Lower level roots of the Slansky algebra}
\begin{center}
\begin{tabular}{|c|c|c|c|}
  \hline
  Form-degree & Positive Root & Field & $\mf{sl}(2,\mathbb{R})$-Representation\\
  \hline
  $0$ & $\alpha_1$ & $\chi$ & \\
  $2$ & $\alpha_0$ & $A^2_2$ & $\mathbf{2}$ \\
  $2$ & $ \alpha_0 + \alpha_1$ & $A^1_2$& $\mathbf{2}$ \\
  $4$ & $2 \alpha_0 + \alpha_1$ & $B_4$ & $\mathbf{1}$\\
  $6$ & $3 \alpha_0 + \alpha_1$ & $\tilde{A}^1_6$ & $\mathbf{2}$ \\
  $6$ & $3 \alpha_0 + 2\alpha_1$ & $\tilde{A}^2_6$ & $\mathbf{2}$ \\
  $8$ & $4 \alpha_0 + \alpha_1$ & $\tilde{\chi}_8$ & $\mathbf{3}$\\
  $8$ & $4 \alpha_0 + 2\alpha_1$ & $\psi_8$ & $\mathbf{3}$\\
  $8$ & $4 \alpha_0 + 3\alpha_1$ & $X_8$ & $\mathbf{3}$\\
  \hline
\end{tabular}
\end{center}
\label{slansky1}
\end{table}
In Table {\bf \ref{slansky1}} we have slightly adapted the notations of \cite{HenryLabordere:2002dk} for the fields associated with the positive roots.

Form-fields coupled to generators with the same $\alpha_0$-level (half the V-degree) have same form degree and  they transform in an irreducible $\mf{sl}(2,\mathbb{R})$ representation. At form degree $0,$ there is the axion $\chi,$ generating the strictly positive part of $\mf{sl}(2,\mathbb{R})$ (adjoint representation $\mathbf{3}$). There is then a doublet of $2$-forms, a single $4$-form which is inert under $\mf{sl}(2,\mathbb{R}),$ a doublet of $6$-forms dual to the $2$-forms and a triplet of $8$-forms.  The field $X_8$ is eliminated by a zero curvature constraint \cite{DallAg:1998} in a supersymmetric and covariant action and the remaining two $8$-forms are dual to the two scalars (axion and dilaton). This is exactly the spectrum of dynamical $p$-forms of type $II_B$ theory, in the duality-invariant formulation. Note that the Slansky algebra encodes in particular the self-duality of the $4$-form, since the $4$-form transforms as the singlet  $\mathbf{1}.$

But the Slansky algebra also contains information about some non-dynamical forms.  Using the 
denominator formula \cite{bor}, one finds at $\alpha_0$-level $5$ the roots given in Table {\bf \ref{slansky2}} 
\cite{Slansky:1992dx,paulot,paulotBis}.
\begin{table}
\caption{Level 5 roots of the Slansky algebra}
\begin{center}\begin{tabular}{|c|c|c|c|}
  \hline
  Form degree & Positive Root & Degeneracy & $sl(2,\mathbb{R})$-Representation(s)\\
  \hline
  $10$ & $5 \alpha_0 + \alpha_1$ & $1$ & $\mathbf{4}$\\
  $10$ & $5 \alpha_0 + 2 \alpha_1$ & $2$ & $\mathbf{4}, \mathbf{2}$ \\
  $10$ & $ 5 \alpha_0 + 3 \alpha_1$ & $2$& $\mathbf{4}, \mathbf{2}$ \\
  $10$ & $5 \alpha_0 + 4 \alpha_1$ & $1$ & $\mathbf{4}$\\
  \hline
\end{tabular} \end{center}
\label{slansky2}
\end{table}
This yields precisely the maximal spectrum of supergravity $10$-forms found subsequenty by a supersymmetry argument in \cite{Bergshoeff:2005ac} -- revised in 
\cite{Bergshoeff:2010hhor}.  Hence the Slansky Borcherds 
algebra with Cartan matrix (\ref{IIB}) remarkably encompasses in a succinct way the complete 
$p$-form spectrum of type $II_B$ supergravity including the non-dynamical forms.

What is true for type $II_B$ is also true for type $II_A.$ The Borcherds superalgebra $\mf{V}_{10_A}$ has in that case the Cartan matrix \cite{HenryLabordere:2002dk}
\begin{equation}A_{II_A} = \left(
     \begin{array}{cc}
       0 & -1 \\
       -1 & 0 \\
     \end{array} 
   \right) \label{IIA} .
\end{equation} 
The first simple root is fermionic now and its root-vector, the corresponding raising operator, is a ``$1$-form", while the second simple root is bosonic with root-vector a ``$2$-form"  (we shall for conciseness but abusively call p-form a generator of V-degree p that couples to a p-form). Both simple roots are null for the given Cartan matrix (but note that on the projective plane blown up at one point, the D0 curve corresponding to the fermionic simple root has self-intersection $-1,$ not $0,$ which suggests that the complete connection to del Pezzo surfaces discussed in the next paragraph  requires in that case a new and broader framework to hold). Using the denominator formula for Borcherds superalgebras \cite{Ray95,Superden}, one finds that the spectrum predicted by the Borcherds superalgebra $\mf{V}_{10_A}$ matches exactly not only the dynamical $p$-forms of type $II_A$ supergravity \cite{HenryLabordere:2002dk}, but also the non-dynamical forms \cite{paulot,paulotBis}, namely the $9$-form and the two $10$-forms, obtained by a supersymmetry argument \cite{Bergshoeff:1996ui,Bergshoeff:2001pv,Bergshoeff:2006qw,Bergshoeff:2010hhor}.

Similar results hold in fact all the way down to three spacetime dimensions, the Cartan matrices of the corresponding Borcherds superalgebras were explicitly worked out in all these cases in \cite{HenryLabordere:2002dk}.  In that paper a correspondence with the middle cohomology of the del Pezzo surfaces was discussed. In the best case of $\mathbb{CP}^1\times \mathbb{CP}^1,$  the $II_B$ Cartan matrix of eq. (\ref{IIB}) turns out to be precisely minus  the intersection form of $2$-cycles. The matching of the supergravity spectrum with the spectrum predicted by the Borcherds superalgebra was verified in \cite{HenryLabordere:2002dk} for the dynamical forms in each case. In fact it remains true for non propagating forms, which among other things allows us to use the denominator formula to obtain the correct roots corresponding to $\pa{D\! - \! 1}$- and $D$-forms.

But the agreement between Borcherds/BKM predictions for both  propagating and  non propagating forms with those of  $E_{11}$ deserves an explanation. It will be seen actually as a consequence of the "extraction" of the former from  $E_{11}$ that is explained in this paper. In fact both methods agree with higher forms permitted by deformation (gauging or mass addition) and supersymmetry arguments, see for instance \cite{Ber07,RW2008,Samt08,Ber09}, resp. \cite{Bergshoeff:1996ui,Bergshoeff:2001pv,Bergshoeff:2006qw,Bergshoeff:2010hhor}.\footnote{The supersymmetry requirements of \cite{Bergshoeff:2005ac} may allow a priori more $D$-forms (top forms) than required by the gauging method \cite{Samt08} that also leads to some of these higher forms.
 \label{caution}}

\subsection{$p$-forms from $E_{11}$}

More recently, different methods based on $E_{11}$ have been used to construct the $p$-form spectra of maximal supergravities. In fact
$E_{11}$ was first considered as a possible spectrum organizing algebra for maximal supergravity theories \cite{julia198}. It is explained there how the representation of the internal symmetry group (the $E_{11-D}$ U-duality) is regularly correlated to the degree of the corresponding set of propagating forms, i.e., to the representation of $ \mf{gl}_D$ for all compactification dimensions $D\ge 3.$ This pointed to branching laws for $E_{11}$ representations into representations of
$\pa{E_{11-D}  \bigoplus   \mf{sl}_D} \bigoplus  \mathbb{R}$
as indicated by the Figure 3 of \cite{julia198} reproduced here as Fig. {\bf 1}, where the star stands for the abelian factor $\mathbb{R}$ coming from the $E_{11}$ generator at root number $D,$ and for $D\le 8$  $E_{11-D}$ is defined by the Dynkin diagram to the right of the star in Fig. {\bf 1}. 

This will be made more precise below. In particular, the reason why one needs a second star for the cases  $E_{1}^A= {\mathbb{R}}$  and $E_{2}=A_1\bigoplus {\mathbb{R}}$ will emerge, as well as the special features of the  cases  $D=10_B$ ($E_{1}^B=A_1$) and $D=2.$

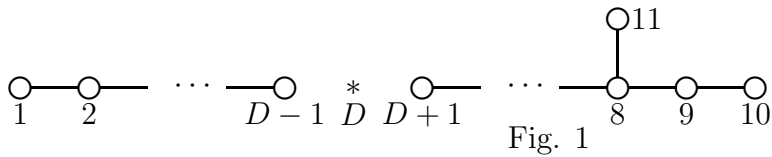
\begin{picture}(190,40)(-4,-22)
\thicklines
\multiput(0,-20)(26,0){2}{\circle{8}}
\put(4,-20){\line(1,0){18}}
\put(30,-20){\line(1,0){18}}
\put(78,-20){\line(1,0){18}}
\put(100,-20){\circle{8}}
\put(152,-20){\circle{8}}
\multiput(226,-20)(26,0){3}{\circle{8}}
\put(0,-30){\makebox(0,0){$1$}}
\put(26,-30){\makebox(0,0){$2$}}
\put(100,-30){\makebox(0,0){$D-1$}}
\put(126,-30){\makebox(0,0){$D$}}
\put(152,-30){\makebox(0,0){$D+1$}}
\put(226,-30){\makebox(0,0){$8$}}
\put(252,-30){\makebox(0,0){$9$}}
\put(278,-30){\makebox(0,0){$10$}}
\put(226,6){\circle{8}}
\put(126,-20){\makebox(0,0){$\ast$}}
\put(237,6){\makebox(0,0){$11$}}
\put(58,-22){$\cdots$}
\put(184,-22){$\cdots$}
\put(156,-20){\line(1,0){18}}
\put(204,-20){\line(1,0){18}}
\put(230,-20){\line(1,0){18}}
\put(256,-20){\line(1,0){18}}
\put(226,-16){\line(0,1){18}}
\put(200,-40){\makebox(0,0){Fig. 1}}
\end{picture}

\bigskip

\bigskip

\bigskip

The subsequent work of \cite{west101} showed that the adjoint representation of $E_{11}$ was the right choice to explain the observations of \cite{julia198}, which considered only the $p$-form spectrum and not the gravitational sector. Reference \cite{west101} most interestingly exhibited a linear-dual graviton as well as the expected 3-form, 6-form and graviton. The present work can be described as the first step towards an interpolation between Borcherds symmetries which leave the graviton inert on the one hand and $E_{11}$ spectrum generating subalgebras that may truly act in the gravity sector but are not understood yet as symmetries on the other.

The level expansion of the adjoint representation of $E_{11}$ according to the number of times the exceptional simple root $\alpha_{11}$ occurs in each root (i.e., its coefficient) yields fields that transform tensorially under $\mf{gl}_{11}$ \cite{Damour:2002cu,West:2002jj}  in $D=11.$  The graviton, the $3$-form, the $6$-form and the linear-dual graviton are located at  ($\alpha_{11}$-)levels $0,$ $1,$ $2$ and $3,$ respectively.   At levels $\geq 3,$ the index-symmetry of the fields are characterized by Young diagrams that  may contain more than one column. For instance, at level $3$ there is the dual graviton $A_{8,1}$ (where the numbers refer to the number of boxes in the successive columns in the Young diagrams), at level $4$ there is an $A_{9,3},$ an $A_{10,1,1}$ and an $A_{11,1}$ and there is a vertiginous exponential explosion of the number of fields with the level \cite{Nicolai:2003fw}.

The fields at levels $\geq 4$ are poorly understood.  An important exception was uncovered a few years ago \cite{Schnakenburg:2002xx,Kleinschmidt:2003mf,Damour:2002fz}, with the identification of a level $4$ field with the massive deformation of type $II_A$ supergravity \cite{Romans:1985tz}:  the type $II_A$ $9$-form descends from the $A_{10,1,1}$ tensor of M theory\footnote{This identification of a level $4$ field with the type $II_A$ massive deformation covers also the coupling to fermions \cite{Henneaux:2008nr}.}.  This was followed more recently by a systematic study of the $p$-form spectrum implied by $E_{11}$ upon dimensional reduction \cite{Riccioni:2007au,Bergshoeff:2007qi}.

The idea is that, when going from $11$ dimensions to $D$-dimensions, some of the fields at higher levels yield $p$-forms. Indeed, the $p$-forms in $D$ dimensions can come not just from the $3$- and $6$-forms in $11$ dimensions, but also from tensors with mixed symmetry when the Young diagram boxes are appropriately saturated with internal indices.  In the latter case one obtains non propagating forms. For instance, for type $II_A,$ one finds the $9$-form mentioned above from $A_{10,1,1},$ and the two $10$-forms from $A_{10,1,1}$ and $A_{11,1}$ \cite{Riccioni:2007au,Bergshoeff:2007qi}. For each dimension, one can derive the spectrum of $p$-forms from the knowledge of the spectrum of $E_{11},$ with the remarkable finding that it agrees with other approaches \cite{Riccioni:2007au,Bergshoeff:2007qi} (but see footnote {\bf\ref{caution}}).

\subsection{Purpose and outline of this paper}

We have thus the following "embarras de richesses".  On the one hand, one can derive the $p$-form spectrum of maximal supergravities by decomposing the Lorentzian Kac-Moody algebra $E_{11}.$
In fact one uses only a parabolic truncation of $E_{11}$ which is  maximal parabolic when there is only one starred node in Fig. {\bf 1}.  This infinite-dimensional algebra contains a huge number of fields of which only a tiny (finite for $D\geq 3$) subset leads to $p$-forms.  On the other hand, one can also derive the $p$-form spectrum from a Borcherds superalgebra, the Cartan matrix of which depends on the spacetime dimension $D.$ In fact one uses essentially only its Borel subalgebra. The BKM superalgebra contains much less excess luggage than $E_{11}$ in the sense that it involves only $p$-forms, and no wider Young diagram of  $ \mf{gl}_D.$  Furthermore the truncation on the Borcherds side is automatic, it is made on the V-degree of the generators through the form degree of the fields that multiply them  as the latter cannot exceed the spacetime dimension. One may also engineer  a truncation at degree $D\! -\! 2$ to drop non propagating fields.

A natural question to be asked is: can one understand why the BKM and $E_{11}$ approaches agree? For instance could one derive the relevant Borcherds algebra, which captures just the $p$-form content of the theory, from $E_{11}$ which contains an infinite number of extra (ill-understood) fields?  The purpose of this paper is to demonstrate that the answer to this question is positive: we shall show how to derive in each spacetime dimension $D \geq 3$ the relevant Borcherds superalgebra from $E_{11}$ \cite{HenryLabordere:2002dk}.  We shall then consider the cases $D=2$ and $D=1$ and we shall prove that $E_{11}$ yields truncations of Borcherds superalgebras in those dimensions as well (although both cases have interesting new features).  These are also new results.

Although our analysis raises many questions, we shall focus here only on the demonstration of the equivalence of the Borcherds and $E_{11}$ approaches in what concerns the $p$-form spectra of maximal supergravities.  We should also refer to \cite{Ber07} for an attempt to truncate $E_{11}$ to a Lie algebra of $p$-forms only; as will be shown in \cite{HJLFuture} there is a Lie algebra quotient of $E_{11}$ at work, for more on this see subsection {\bf 4.1}. The present superalgebra approach does a similar truncation efficiently. Comments on some open problems will be given with the conclusions.

Our paper is organized as follows.  
In the next section (section {\bf \ref{covariant}}), we give, for each $D,$ a presentation of $E_{11}$ which is $\mf{gl}_D\pa{\mathbb{R}}$-covariant with a subtlety in the case $D=2$.  
We then tensorize a parabolic (definition is recalled in subsection {\bf 2.4}) subalgebra of $E_{11}$ that contains 
\begin{equation}
\mf{gl}_D\pa{\mathbb{R}} \oplus E_{11-D} \label{SUM}
\end{equation}
by the Grassmann algebra in $D$ dimensions (section {\bf \ref{tensoring}}) and investigate the structure of the  sub-superalgebra of $\mf{gl}_D\pa{\mathbb{R}}$-invariants (section {\bf \ref{pformalgebra}}). This algebra is just a V-duality superalgebra of symmetries 
of $p$-forms in $D$ dimensions and we can describe it in terms of generators and relations.  
We then verify in section {\bf \ref{idBorcherds}} that for $D\ge 3$ these generators and relations are precisely those of the Borcherds superalgebras of \cite{HenryLabordere:2002dk}, which completes the proof of the equivalence. Section   {\bf \ref{LOW}} discusses the peculiarities of the  low dimensional cases $D=2$ and $D=1$. 
Our last section is devoted to concluding comments.

\section{Covariant presentations of $E_{11}$}
\label{covariant}
\setcounter{equation}{0}
\subsection{$\mf{gl}_{11}\pa{\mathbb{R}}$ inside $E_{11}$}
The Chevalley-Serre generators of $E_{11}$ are denoted by $h_i,$ $e_i$ and $f_i$ ($i=1, \cdots, 11$).  If one removes the exceptional node numbered $11$ in Fig. {\bf 1}  above, one remains with the Dynkin diagram $A_{10}$ of $\mf{sl}_{11}\pa{\mathbb{R}},$ which is a regular subalgebra of $E_{11}$ (its Cartan subalgebra  is included in that of $E_{11}$).  $A_{10}$ defines the so-called ``gravity line" \cite{Nuf80} in $11$ dimensions.  It is common usage to denote the $e_i$'s associated with the gravity line $K^i_{\; i +1}$ ($i = 1, \cdots, 10$) and we shall follow that practice here.  The corresponding $f_i$'s are $K^{i+1}_{\; \; \; \;  i}$ ($i = 1, \cdots, 10$). The Cartan generators $h_i$ ($i = 1, \cdots, 10$) must be rewritten as  $h_i =  K^i_{\; i} - K^{i+1}_{\; \; \; \; i +1}$ ($i = 1, \cdots, 10$).  The root vector $e_{11}$ associated with the exceptional node is denoted $E^{9 \ts 10 \ts 11}$ \cite{LuPope96}.

It is well known that by using the Cartan generator $h_{11}$ associated with the exceptional node, one can extend the embedding of $\mf{sl}_{11}\pa{\mathbb{R}}$ in $E_{11}$ to a regular embedding of $\mf{gl}_{11}\pa{\mathbb{R}}$ in $E_{11}.$  For that purpose, one needs to express the central generator $K_{11}=K^1_{\; 1}+K^2_{\; 2}...+K^{10}_{\; \;10}+K^{11}_{\; \; 11}$ of $\mf{gl}_{11}\pa{\mathbb{R}}$ in terms of the Cartan generators $h_i.$  There are many ways to do so. The reason is that, while $K_{11}$ should commute of course with the generators of $\mf{sl}_{11}\pa{\mathbb{R}},$ its commutation relation with the root-vector $E^{9 \ts 10 \ts 11}$ is conventional if we only require that  the embedding be regular.  Different choices correspond to assigning different density weights to $E^{9 \ts 10 \ts 11}.$  If one requests that $E^{9 \ts 10 \ts 11}$ be the ($9$-$10$-$11$)-th component of a $3$-form with no extra density weight, one must impose $[K_{11},E^{9 \ts 10 \ts 11}] = 3 E^{9 \ts 10 \ts 11}.$ This convenient choice fixes completely $K_{11}$ and will be adopted here.  Explicitly one finds
\begin{equation}
K_{11}=-\frac{3}{2} [3\pa{h_1+2h_2 + \cdots +8h_8} + 8\pa{ h_{10} +2h_9  } +11 h_{11}].  \label{K11}
\end{equation}
The first parenthesis on the right-hand side can be recognized as a Cartan generator that extends  $\mf{sl}_{8}\pa{\mathbb{R}}$ to  $\mf{gl}_{8}\pa{\mathbb{R}}$ inside $\mf{sl}_{9}\pa{\mathbb{R}}$ .

Knowing $K_{11}$ defines completely the embedding of $\mf{gl}_{11}\pa{\mathbb{R}}$ in $E_{11}.$  All the basis elements $K^\gamma_{\; \; \lambda}$ of $\mf{gl}_{11}\pa{\mathbb{R}}$ ($\gamma, \lambda = 1, \cdots, 11$) can be expressed in terms of the Chevalley-Serre generators $h_i$ ($i=1, \cdots, 11$) and $e_i,$ $f_i$ ($i=1, \cdots, 10$) of $E_{11}$ and their multiple commutators.  They fulfill the algebra
\begin{equation}
 [K^\gamma_{\; \; \lambda},K^\alpha_{\; \; \xi}]=\delta^\alpha_{\; \; \lambda} K^\gamma_{\; \; \xi} -\delta^\gamma_{\; \; \xi} K^\alpha_{\; \; \lambda}.
\end{equation}

One can conversely express the Cartan generator $h_{11}$ in terms of the Cartan generators of $\mf{gl}_{11}\pa{\mathbb{R}},$
\begin{equation} h_{11}=-\frac{1}{3}K_{11}+\pa{K_{\; 9}^9+K_{\;10}^{10}+K_{\;11}^{11}} . \end{equation}

\subsection{Gravity line $A_{D-1}$ and embedding of $\mf{gl}_D\pa{\mathbb{R}}$ in $E_{11}$}

We now proceed with the dimensional reduction to $D$ spacetime dimensions. We begin by choosing, for all dimensions $D$ except for the case $10_B,$ the node numbered $D$ by marking it with a star, as in Fig. {\bf 1}. {}For the case $10_B,$ we mark the node numbered $9.$ In the case $10_B,$ it is convenient to change the numbering of the Dynkin nodes and to use primed indices.  The starred node is then $10'.$ The internal group is associated with the node $11'.$ The $10_B$ Dynkin diagram reads now:
\bigskip

\begin{picture}(190,40)(-20,-22)
\thicklines
\multiput(0,-20)(26,0){9}{\circle{8}}
\put(0,-30){\makebox(0,0){$1'$}}
\put(26,-30){\makebox(0,0){$2'$}}
\put(52,-30){\makebox(0,0){$3'$}}
\put(78,-30){\makebox(0,0){$4'$}}
\put(104,-30){\makebox(0,0){$5'$}}
\put(130,-30){\makebox(0,0){$6'$}}
\put(156,-30){\makebox(0,0){$7'$}}
\put(182,-30){\makebox(0,0){$8'$}}
\put(208,-30){\makebox(0,0){$9'$}}
\put(182,0){\makebox(0,0){$\ast$}}
\put(193,0){\makebox(0,0){$10'$}}
\put(182,20){\circle{8}}
\put(193,20){\makebox(0,0){$11'$}}
\put(4,-20){\line(1,0){18}}
\put(30,-20){\line(1,0){18}}
\put(56,-20){\line(1,0){18}}
\put(82,-20){\line(1,0){18}}
\put(108,-20){\line(1,0){18}}
\put(134,-20){\line(1,0){18}}
\put(160,-20){\line(1,0){18}}
\put(186,-20){\line(1,0){18}}
\put(182,-50){\makebox(0,0){Fig. 2}}
\put(182,-70){\makebox(0,0){}}
\end{picture}
 
  \bigskip
  \bigskip
 
  \bigskip
  
What stands to the left of the marked node (actually, below the marked node for $D=11$ or for 
$D=10_B$) is a Dynkin diagram of type $A_{D-1},$ except in two cases. \begin{itemize}
\item For $D=10_A$ the diagram is of type $D_{10}$ and we also mark the node numbered $11$ to get a remaining $A_9.$
\item For $D=9,$ the diagram is of type $A_9$ (not $A_8$), and we mark again as second node, the node numbered $11$ (see Fig. {\bf 3}).
\begin{center}
\begin{picture}(190,40)(-4,-22)
\thicklines
\multiput(0,-20)(26,0){2}{\circle{8}}
\put(4,-20){\line(1,0){18}}
\put(30,-20){\line(1,0){18}}
\put(58,-22){$\cdots$}
\put(78,-20){\line(1,0){18}}
\put(100,-20){\circle{8}}
\put(100,-30){\makebox(0,0){$8$}}
\put(111,6){\makebox(0,0){$11$}}
\put(100,6){\makebox(0,0){$\ast$}}
\put(126,-20){\makebox(0,0){$\ast$}}
\put(152,-20){\circle{8}}
\put(0,-30){\makebox(0,0){$1$}}
\put(26,-30){\makebox(0,0){$2$}}
\put(126,-30){\makebox(0,0){$9$}}
\put(152,-30){\makebox(0,0){$10$}}
\put(100,-50){\makebox(0,0){Fig. 3}}
\put(100,-70){\makebox(0,0){}}

\end{picture}
\end{center}

\bigskip

\bigskip

What remains to the left of the marked nodes is $A_8.$
\end{itemize}

\bigskip

When this is done, we are left with a Dynkin subdiagram of type $A_{D-1}$ in all cases. This $A_{D-1}$ defines the so-called ``gravity line".  The corresponding subalgebra $\mf{sl}_D\pa{\mathbb{R}}$ regularly embedded in $E_{11}$ is generated by the first $D\! -\! 1$ generators $h_i,$ $e_i$ and $f_i$ ($i=1, \cdots, D\! -\! 1$) for $D \! \not= \! 10_B.$ For $D \! = \! 10_B,$ the gravity line $A'_9$ is defined by the nodes numbered $1$ to $8$ and the node numbered $11$ (which becomes $9'$ after the renumbering described above).

We can extend the embedding of $\mf{sl}_D\pa{\mathbb{R}}$ to an embedding of $\mf{gl}_D\pa{\mathbb{R}}$ in $E_{11}$ by simply taking $\mf{gl}_D\pa{\mathbb{R}}$ to lie inside the above $\mf{gl}_{11}\pa{\mathbb{R}}$ in the natural way. This is equivalent to requiring that all relevant tensors carry no density weight.  We denote the standard basis of $\mf{gl}_D\pa{\mathbb{R}}$ by $\{K^\al_{\; \; \be}\}$ ($\al, \be = 1, \ldots, D$) and let us define the diagonal generator $K_D$ by $K_D:=\sum_{\al = 1}^{D} K^\al_{\; \; \al}.$  In every dimension $D$ the $K^\al_{\; \; \be}$'s, and in particular $K_D,$  are given in terms of the Chevalley-Serre generators of $E_{11}$ by the same expressions as in $11$ dimensions.  Detailed formulas are worked out in \cite{HJLFuture}.

The only exception to this discussion is type $II_B$ since the gravity line contains in that case the exceptional node and is not a subline of the gravity line in $11$ dimensions. However, there is again a natural extension of the $A'_9$ defined by the $II_B$-gravity line to $\mf{gl}'_{10}\pa{\mathbb{R}}.$  It is obtained by demanding that the root-vector $e_{9}$ transforms as the ($9'$-$10'$)-th component of a $\mf{gl}'_{10}\pa{\mathbb{R}}$ $2$-form without extra density weight (recall that it is attached to the second node of the $II_B$-gravity line) and that the root-vector $e_{10}$ be a true scalar of $\mf{gl}'_{10}\pa{\mathbb{R}}$ (and not a scalar density with non-trivial weight).  The corresponding expression is easily worked out and 
reads $$K_{10_B}= -2[2\pa{h_{1'}+2h_{2'} + \cdots +8h_{8'}} + 8h_{9'}  +10 h_{10'}+5 h_{11'}].$$

\subsection{U-duality algebra $E_{11-D}$}

The unmarked nodes (with no star) which are not on the gravity line define the semi-simple part of the duality subalgebra $E_{11-D}.$  It turns out that the two cases with two stars correspond precisely to situations where the gravity line is adjacent to more than one node of $E_{11},$ in fact to the two starred  nodes. Each of these special cases is a one dimensional reduction of a maximal dimension theory with $D=10$ for $10_B$  giving $D=9,$ resp. $D=11$ reducing to $10_A.$ 

\subsubsection{Case of one starred node}

$E_{11-D}$  is semi-simple precisely when there is only one starred node. The Chevalley-Serre generators of $E_{11}$ split into the Chevalley-Serre generators of $A_{D-1},$ the Chevalley-Serre generators of $E_{11-D}$ and the Chevalley-Serre generators $\{h_D, e_D, f_D\}$ ($\{h_9, e_9, f_9\}$ for $10_B$) associated with the starred node.  In order to study the $\mf{gl}_D\pa{\mathbb{R}}$-invariants, it is convenient to trade the Chevalley-Serre generator $h_D$ with $K_{11},$ which commutes with the $\mf{gl}_D\pa{\mathbb{R}}$ subalgebra.  This can be done for all $D$'s since the coefficient of $h_D$ in the expansion of $K_{11}$ is never zero (see (\ref{K11})).

Alternatively, for all dimensions except $D=2,$ one can replace the Cartan generator $h_D$ by $K_D$ exhibiting thereby a $\mf{gl}_D\pa{\mathbb{R}} \oplus E_{11-D}$ subalgebra. The fact that  $\mf{sl}_D\pa{\mathbb{R}}\oplus E_{11-D}$ is a direct sum is (even for D=2) a general property of regular subalgebras. The reason why this cannot be extended to $\mf{gl}_D$  in $D=2$ dimensions is that the trace $K_2$ as defined above does not involve $h_2,$  it is a linear combination of the Cartan generators of $E_9.$  $K_2$ coincides in fact with the central charge of $E_9$ \cite{BJH81,HJLFuture}.  In other dimensions there is no central generator of $E_{11-D}$, it would have commuted with $\mf{sl}_D,$.

Since it will be important in the sequel to have the full $\mf{gl}_D\pa{\mathbb{R}}$-symmetry  manifest, we shall take in the $D=2$ case a basis of Cartan generators that contains all the Cartan generators of $\mf{gl}_2\pa{\mathbb{R}}$ including $K_2,$  $K_{11}$ (to replace $h_2$) as well as 8 Cartan generators among $h_3, \cdots, h_{11}$ that span a complementary space.  Other choices avoiding $K_{11}$ are possible
and in fact necessary to exhibit again on the Borcherds side a nice presentation with all the information contained in the Cartan matrix, this will be explained in \cite{HJLFuture} . 

\subsubsection{Case of two starred nodes}
The case of two starred nodes corresponds to a non semisimple U-duality subalgebra $E_{11-D},$ which is then the semi-simple Lie algebra defined by the unmarked nodes which are not on the gravity line, plus an extra $\mathbb{R}$-factor. The generator of this extra $\mathbb{R}$-factor can be taken to be $h_*=h_*^{10_A}=K^{11}_{\; \; 11}$
 in $D=10_A$ dimensions, and $h_*=h_*^9=K^{10}_{\; \; 10}+K^{11}_{\; \; 11}$
 in $D=9$ dimensions. These generators $h_*$  are $\mf{gl}_D\pa{\mathbb{R}}$-scalars.

A basis of the Cartan subalgebra of $E_{11}$ adapted to those cases is given by the $10$ Cartan generators of $\mf{gl}_{10}\pa{\mathbb{R}}$ together with $h_*^{10_A}$ in the case $D=10_A,$ and the $9$ Cartan generators of $\mf{gl}_{9}\pa{\mathbb{R}}$ together with $h_*^9$ and $K^{10}_{\; \; 10} - K^{11}_{\; \; 11}$ (the Cartan generator of the unmarked -- internal $A_1$ -- node not on the gravity line)  in the case $D=9.$

\subsection{Parabolic subalgebra $\mathcal{P}E_{11}\pa{D}$}
By definition a parabolic subalgebra of a BKM algebra is any subalgebra that contains a Borel subalgebra (the ``upper-triangular subalgebra" in the case of $\mf{gl}_{n}\pa{\mathbb{R}}$), i.e., all the  Cartan diagonal generators ($h_i$) and all the ``raising" operators generated by the $e_i$ simple positive generators of it. It turns out that in the case of a  finite dimensional  semi-simple complex  Lie algebra it can be obtained up to conjugation by adding to any given Borel subalgebra a subset of the simple negative generators $f_i$ (the simple ``lowering" operators) and their commutators. A maximal parabolic proper subalgebra generating set misses only one generator $f_i$ among all the generators of the presentation.

We consider for each $D$ the smallest parabolic subalgebra of $E_{11}$ containing the subalgebra $\mf{sl}_D \bigoplus E_{11-D},$ which we will call $\mathcal{P}E_{11}\pa{D}.$ This subalgebra is generated by all the $h_i$'s, all the $e_i$'s as well as those $f_i$'s  that are not associated with the starred node(s).  Hence it is a maximal parabolic subalgebra when there is only one starred node.  It is not maximal parabolic in the other cases $D=10_A$ and $D=9.$

We shall from now on focus on the parabolic subalgebras $\mathcal{P}E_{11}\pa{D}.$  We shall show how to construct from each $\mathcal{P}E_{11}\pa{D}$ a corresponding parabolic subalgebra of a Borcherds superalgebra.  Once the parabolic subalgebra of the Borcherds superalgebra is determined, the full Borcherds superalgebra is in fact known.  The missing lowering operators are simply defined by symmetry ie using the Chevalley involution.  This procedure yields in particular  the Borcherds superalgebras of \cite{HenryLabordere:2002dk} for $D\ge 3.$

\subsection{Fundamental forms in D dimensions}
\label{dec}

It is clear that the adjoint representation of $E_{11}$ decomposes into representations of $\mf{gl}_D.$ Among these, one has the adjoint representation described by the generators $K^\alpha_{\; \beta},$  $\alpha , \beta =1, \ldots , D.$  We have also seen that a basis of Cartan generators of $E_{11}$ not in $\mf{gl}_D$ {\it can} be taken to be $\mf{gl}_D$-scalars.

We investigate in this subsection the $\mf{gl}_D$-representations into which the root-vectors $e_i$ not associated with the gravity line transform. As they turn out to all transform as $p$-forms for some $p$ (including $p=0$), we call these representations the ``fundamental form" representations.

It is clear that the non-gravity root-vectors $e_i$ associated with the unmarked nodes are scalars, so let us now turn to the ones associated with the starred node(s).
For all spacetime dimensions, each starred node in the diagram of $E_{11}$ is attached to one node of the gravity line via a single link, the $p^{th}$ node say, starting from the right of the gravity line. This implies that the corresponding Chevalley-Serre raising operator is the lowest weight state of the fundamental representation of $\mf{gl}_D\pa{\mathbb{R}}$ with Dynkin labels
$(0,0, \ldots, -1, \ldots, 0)$ (only one $-1,$ in position $p$).  This is the ``$p$-form" representation (antisymmetric tensors of rank $p$ by a consistent choice of conventions). Hence, we conclude that under the action of $\mf{gl}_D\pa{\mathbb{R}}$ the simple roots of $E_{11}$ not on the gravity line generate indeed only $p$-form representations including scalars, as we announced above. The list of the fundamental representations that appear for each $D$ is given in table \ref{Table1}.
\begin{table}
\begin{center}
\begin{tabular}{|c||c|}
\hline \rule[-1mm]{0mm}{6mm}
D & Fundamental form/  lowest weight state\\
\hline \rule[-1mm]{0mm}{6mm}
11  & $3$-form $E^{\al \be \ga}$/ $E^{9\ts 10 \ts 11}$\\
\hline \rule[-1mm]{0mm}{6mm}
$10_A$ ($II_A$) & $1$-form $K^\al$/ $K^{10}_{\; \; (11)}$ ; $2$-form $E^{\al \be}$/ $E^{9\ts 10 \ts (11)}$\\
$10_B$ ($II_B$) & scalar  $T^+$/ $K^{(10)}_{\; \;  \; (11)}$ ; $2$-form $E'^{\al' \be'}$/ $E'^{9'\ts 10'} = K^{9}_{\; 10}$\\
\hline \rule[-1mm]{0mm}{6mm}
9 & scalar $T^+$/ $K^{(10)}_{\; \;  \; (11)}$ ; two $1$-forms $K^\al$/ $K^9_{\; (10)}$ and $E^\al$/ $E^{9\ts (10) \ts (11)}$ \\
\hline \rule[-1mm]{0mm}{6mm}
$1\leq D \leq 8$ & $11\! - \! D$ scalars (axions) of $E_{11-D}$ ; $1$-form $K^\al$/ $K^D_{\; (D+1)}$\\
\hline
\end{tabular}
\end{center}
\caption{\small This table gives for all fundamental forms in $D$ dimensions their $\mf{gl}_D\pa{\mathbb{R}}$ lowest weight states.  Internal indices are within parentheses. Scalars ($0$-forms) have been included. The greek indices run from $1$ to $D.$ \label{Table1}}
\end{table}

We shall see that each of these fundamental $p$-form generators will become a raising Chevalley generator of the relevant Borcherds superalgebra.

\subsection{$\mf{gl}_D\pa{\mathbb{R}}$-covariantization of the Chevalley-Serre relations}
\label{covariantizationZZZ}

Our next step is to provide a presentation of $E_{11}$ which is manifestly covariant with respect to $\mf{gl}_D \pa{\mathbb{R}},$ for any $D.$

\vspace{.2cm}
\noindent
{\bf Definition}: A set of generators and relations involving $E_{11}$-elements is called a $\mf{gl}_D\pa{\mathbb{R}}$-covariant presentation of $E_{11}$ if and only if \begin{enumerate} \item The generators and relations span complete representations of $\mf{gl}_D\pa{\mathbb{R}}.$  \item The relations are consequences of the standard $E_{11}$-Chevalley-Serre relations. \item The standard $E_{11}$-Chevalley-Serre relations can conversely be derived from the given set of relations. \end{enumerate}  The covariant presentations are redundant, in the sense that some of the relations contained in the set are consequences of the others.  {}Furthermore, there exist various $\mf{gl}_D\pa{\mathbb{R}}$-covariant presentations of $E_{11},$ as a $\mf{gl}_{D+1}\pa{\mathbb{R}}$-covariant presentation is automatically $\mf{gl}_D\pa{\mathbb{R}}$-covariant.  We shall construct here ``minimal" covariant presentations.

The standard Chevalley-Serre presentation of $E_{11}$ is not manifestly $\mf{gl}_D\pa{\mathbb{R}}$-covariant as it involves generators of the algebra that do not span complete representations of $\mf{gl}_D\pa{\mathbb{R}}.$  However, since $E_{11}$ contains $\mf{gl}_D\pa{\mathbb{R}},$ it is guaranteed that one can covariantize the Chevalley-Serre relations in order to get a $\mf{gl}_D\pa{\mathbb{R}}$-covariant presentation. The simplest solution is the ``maximal" presentation obtained by choosing a full linear basis of the Lie algebra as generators and all commutation relations as relations. But an explicit ``minimal" $\mf{gl}_D\pa{\mathbb{R}}$-covariantization of  the Chevalley-Serre presentation is in fact straightforward to achieve. We shall return \cite{Future} to general results on covariant presentations of an algebra with respect to a subalgebra.  Here, we shall just list the covariant presentations of $E_{11}$ with respect to $\mf{gl}_D \pa{\mathbb{R}}$ for all $D$'s, verifying explicitly that the properties of a covariant presentation are indeed fulfilled for $D \! = \! 11,$ and leaving the verification for the other dimensions to the reader.  It is actually sufficient for our purposes to covariantize the parabolic subalgebra $\mathcal{P}E_{11}\pa{D},$ and this is what shall be considered here. It is straightforward to extend the analysis to the full $E_{11}.$

In all cases, the $\mf{gl}_D\pa{\mathbb{R}}$ covariantization of the relations defining $\mf{gl}_D\pa{\mathbb{R}}$ itself simply amounts to consider the entire adjoint representation with basis $K^\al_{\; \be},$ $1 \leq \al, \beta \leq D,$ which obey \begin{equation}[K^\al_{\; \be},K^\ga_{\; \delta}]=\delta^\ga_{\; \be} K^\al_{\; \delta} -\delta^\al_{\; \delta} K^\ga_{\; \be}.\end{equation}  It involves the trivial central extension of  $\mf{sl}_D\pa{\mathbb{R}}$ to $\mf{gl}_D\pa{\mathbb{R}}.$ Similarly, the standard Chevalley-Serre presentation of the internal duality group $E_{11-D}$ is manifestly $\mf{gl}_D\pa{\mathbb{R}}$-covariant since the generators of $E_{11-D}$ are $\mf{gl}_D\pa{\mathbb{R}}$-scalars.  Therefore, one needs to treat only the Chevalley-Serre relations involving the generators associated with the starred nodes.  In fact, when there is a single starred node, one can trade as we have seen the associated Cartan generator for the trace of $\mf{gl}_D\pa{\mathbb{R}}$ (except for $D=2$), and so, the relations involving it are automatically contained in the $\mf{gl}_D\pa{\mathbb{R}}$-covariantized relations. Therefore only the raising operator associated with the starred node needs to be explicitly considered. {}For the case of two starred nodes, or for $D=2,$ there is one additional Cartan generator to be taken into account.

\subsubsection{$D=11$}
The covariantized generators associated with the starred node $11$ are $E^{\lambda \pi \sigma},$ since $E^{9 \ts 10 \ts 11}$ is the lowest weight state of the $3$-form representation. Let us repeat that the term 3-form refers strictly speaking to the contragredient representation, namely that of the fields that multiply the corresponding "3-form" generators. The other components of $E^{\lambda \pi \sigma}$ are obtained through successive commutators of $E^{9 \ts 10 \ts 11}$ with the raising operators $K^\al_{\; \al + 1}$ of $\mf{gl}_D\pa{\mathbb{R}}.$ The covariantized Chevalley-Serre relations involving $E^{\lambda \pi \sigma}$ are
\begin{eqnarray}
\left[K^\lambda_{\; \pi},E^{\theta \phi \psi} \right]&=& 3 \ts \ts \delta^{[\theta |}_{\; \pi} E^{\lambda| \phi \psi]} ,\label{commKE} \\
\left[E^{\lambda \pi \sigma},E^{\theta \phi \psi} \right] &=& \left[E^{[\lambda \pi \sigma},E^{\theta \phi \psi]} \right] \! \! , \label{commEE}
\end{eqnarray}
where antisymmetrization (indicated by bracketing the indices) carries weight one, i.e., is idempotent. The first relation expresses that $E^{\theta \phi \psi}$ transforms as a $3$-form. The second relation expresses that $[E^{\lambda \pi \sigma},E^{\theta \phi \psi}],$ which is a priori in the antisymmetric tensor product of the $(0,0,0,0,0,0,0,-1,0,0)$ representation with itself, contains only the fully antisymmetric part\footnote{The antisymmetric tensor square of the antisymmetric 3-form decomposes into exactly two irreducible representations of $\mf{gl}_{11}\pa{\mathbb{R}}.$ }. 
These relations are well known to be consequences of the Chevalley-Serre relations of $E_{11}.$  

In turn, one checks without difficulty that they imply them. For instance, it follows from (\ref{commEE}) that $0 = [E^{8 \ts 10 \ts 11}, E^{9 \ts 10 \ts 11}] = [[K^8_{\; 9},E^{9 \ts 10 \ts 11}],E^{9 \ts 10 \ts 11}].$ Equation  (\ref{commEE})  is the covariantization of this Serre relation, it expresses the vanishing of the irreducible representation generated by the single component  $[E^{8 \ts 10 \ts 11}, E^{9 \ts 10 \ts 11}].$

In the same way, eq. (\ref{commKE}) implies $0 = [K^8_{\; 9}, E^{8 \ts 10 \ts 11}] = [K^8_{\; 9},[K^8_{\; 9},E^{9 \ts 10 \ts 11}]].$ This second Serre relation results from the antisymmetry of the 3-form. The other Chevalley-Serre relations are easily verified along similar lines.

\subsubsection{$D=10,$ $II_A$}
The covariant generators are $K^\al_{\; \be},$ $h_* = K^{11}_{\; \; 11} ,$ $E^{ \be \ga \ts 11}$ and $K^\al_{\; 11}.$
The covariant form of the Chevalley-Serre relations read \begin{eqnarray} &&
\left[E^{ \al \be \ts 11},E^{ \rho \sigma \ts 11} \right]=0, \; \left[K^\lambda_{\; 11},K^\mu_{\; 11} \right]=0,\\
&&
 \left[K^\lambda_{\; 11}, E^{ \mu \nu \ts 11} \right]= \left[K^{[\lambda}_{\; 11}, E^{ \mu \nu ] \ts 11} \right] \! \! , \\
&& 
\left[K_{\; \;  11}^{11},K^\al_{\; \be} \right]=0, \\ 
&&
\left[K_{\; \; 11}^{11},K^\al_{11}\right]=-K^\al_{\; 11}, \left[K^\al_{\; \be},K^{\gamma}_{\; 11} \right]= \delta^{\ga}_{\; \be} K^\al_{\; 11},
\\ && \left[K_{\; \; 11}^{11},E^{\al \be \ts 11} \right]=  E^{ \al \be \ts 11}, \left[K^\al_{\; \be},E^{ \mu \nu \ts 11} \right]= \delta^{\nu}_{\; \be} E^{ \mu \al \ts 11}
+ \delta^{\mu}_{\; \be} E^{ \al \nu \ts 11}.
\end{eqnarray}
The last line says that $E^{ \mu \nu \ts 11}$ is a $2$-form for $\mf{gl}_{10}$ and the penultimate that $K^\lambda_{\; 11}$ is a $1$-form.

\subsubsection{$D=10,$ $II_B$}
The covariant generators of $\mathcal{P}E_{11}\pa{10_B}$ are $K^\al_{\; \be},$ $E^{\al \be},$ $e=e_{11'},$ $h_{11'},$ $f = f_{11'}.$ The covariantized relations involving the raising generator $E^{\al \be}$ associated with the starred nodes are:
\begin{eqnarray}
&&
\left[ K^\al_{\; \be},E^{\ga \delta} \right] = -2 \ts \delta^{[\ga}_{\; \be} E^{\delta]\al}, \; \left[ E^{\al \be},E^{\ga \delta} \right] = 0,
\\ && \left[ e, \left[ e, E^{\al \be} \right] \right]= 0,
\; \left[h_{11'},E^{\al \be} \right]=-E^{\al \be}, \; \left[ f, E^{\al \be} \right] = 0.
\end{eqnarray} The first relation defines $E^{\al \be}$ as a $2$-form.

\subsubsection{$D=9$}
The covariantized generators of $\mathcal{P}E_{11}\pa{9}$ are $K^\al_{\; \be},$ $E^{\al \ts 10 \ts 11 },$ $K^\al_{\; 10},$ $h_* = K^{10}_{\; \; 10} + K^{11}_{\; \; 11} $ (alternatively, $h'_* = K^{10}_{\; \; 10}$) and $e_{10}, h_{10}, f_{10}.$
Besides the relations that express that $E^{\al \ts 10 \ts 11 }$ and $K^\al_{\; 10}$ are $1$-forms for $\mf{gl}_9,$ and $h_*$ a scalar, and that $h_*$ commutes with the internal $\mf{sl}_2\pa{\mathbb{R}}$ duality algebra,  the covariantized Chevalley-Serre relations involving $E^{\al \ts 10 \ts 11 },$ $K^\al_{\; 10}$ and $h_*$ read:
\begin{eqnarray} &&
\left[E^{\al \ts 10 \ts 11 },E^{\be \ts 10 \ts 11 } \right] = 0, \; \left[K^\al_{\; 10},K^\be_{\; 10} \right] = 0,
 \\ && \left[E^{\al \ts 10 \ts 11 },K^\be_{\; 10} \right] = \left[E^{10\ts 11 \ts [\al},K^{\be]}_{\; 10} \right]\!  \!  , \\ &&
\left[h_*,E^{\al \ts 10 \ts 11 } \right]= 2 E^{\al \ts 10 \ts 11 } , \;
\left[h_*,K^\al_{\; 10} \right]= - K^\al_{\; 10}, \\ &&
\left[e_{10},\left[e_{10}, K^\al_{\; 10}\right] \right] = 0, \,
\left[K^\al_{\; 10},\left[K^\be_{\; 10}, e_{10} \right] \right] = 0, \,
\left[e_{10}, E^{\al \ts 10 \ts 11 } \right]= 0, \\ &&  \left[f_{10},E^{\al \ts 10 \ts 11 } \right]= 0, \; \left[f_{10},K^\al_{\; 10} \right]= 0 , \\
&& \left[h_{10},E^{\al \ts 10 \ts 11 } \right]= 0, \; \left[h_{10},K^\al_{\; 10} \right]= - K^\al_{\; 10}. \end{eqnarray}

\subsubsection{$D\leq 8$}
\hspace{1mm}
The covariantized generators of $\mathcal{P}E_{11}\pa{D}$ are $K^\al_{\; \be},$ $K^\al_{\; D+1}$ and the generators of $E_{11-D}.$  The covariantized Serre-relations split into: \begin{itemize}
\item the commutation relations of $\mf{gl}_D\pa{\mathbb{R}}$;
\item the Chevalley-Serre relations of the U-duality group $E_{11-D}$;
\item the relation \begin{equation} \left[K^\al_{\; \be},K^{\gamma}_{\; D+1} \right]= \delta^{\ga}_{\; \be} K^\al_{\; D+1} \end{equation} that expresses that $K^\al_{\; D+1}$ transforms as a $\mf{gl}_D\pa{\mathbb{R}}$ $1$-form;
\item the relation that describes how $K^\al_{\; D+1}$ transforms under the U-duality group; these are \begin{eqnarray} && \left[e_{D+1},\left[e_{D+1},K^\al_{\; D+1}\right] \right] =0, \, \left[K^\al_{\; D+1},\left[K^\be_{\; D+1}, e_{D+1} \right] \right] =0, \\ && \left[h_{D+1},K^\al_{\; D+1} \right]=-K^\al_{\; D+1}, \; \left[f_{D+1},K^\al_{\; D+1}\right] = 0,
\end{eqnarray} for $D \leq 7,$ and in addition \begin{eqnarray} && \left[e_{11},\left[e_{11},K^\al_{\; 9}\right] \right] =0, \; \left[K^\al_{\; 9},\left[K^\be_{\; 9}, e_{11} \right] \right] =0, \\ && \left[h_{11},K^\al_{\; 9} \right]=-K^\al_{\; 9}, \; \left[f_{11},K^\al_{\; 9
}\right] = 0 ,\end{eqnarray}for $D =8$ (the other commutation relations of $K^\al_{\; D+1}$ with the internal generators are zero);
\item the commutation relation \begin{equation} \left[K^\lambda_{\; D+1},K^\mu_{\;D+1} \right]=0 .\end{equation}
\end{itemize}
In the case $D=2,$ there is an extra Cartan generator as we have seen, which can be taken to be $K_{11}.$  This is a spacetime scalar that commutes with all the other generators except $e_{11}$ and $f_{11},$ for which one has $[K_{11}, e_{11}] = 3 \ts e_{11},$ $[K_{11}, f_{11}] = -3 f_{11}.$

To summarize: 
in all cases but $D=2,9,10_A,$ the generators of the covariant presentation of $\mathcal{P}E_{11}\pa{D}$ are
$$K^\al_{\; \be} \, \, (1 \leq \al,\be \leq D), \, A^{\textbf{cov}}, \, e_a, \, f_a, \, h_a,$$ where $a$ indexes  an internal $U$-duality's Cartan subalgebra basis and where
 $A^{\textbf{cov}}$ stands for the covariantized raising generators associated with the starred Dynkin node(s) ($3$-form, $2$-form, $1$-form).
In the remaining cases $D=9,10_A,$ one must complete the set of generators by adding for instance the extra Cartan element that we called $h*$ and which is a spacetime scalar. We could have taken instead $K_{11}$ as we have explained above, or any other convenient linearly independent Cartan element.  Finally, for $D=2,$ one must also add one Cartan element which again may be taken to be $K_{11}$ but one recalls that the $h_a$'s and the diagonal generators $K^\al_{\; \al}$ are not linearly independent in that case.


\section{Tensoring with the exterior algebra $\Lambda\! \pa{\mathbb{R}^{D}}$}
\label{tensoring}
\setcounter{equation}{0}

We now take the tensor product of the purely bosonic Lie algebra $\mathcal{P}E_{11}\pa{D}$ with the Grassmann superalgebra $\Lambda \!\pa{\mathbb{R}^{D}}$ constructed on a $D$-dimensional vector space\footnote{This is just the exterior algebra with the $\mathbb{Z}_2$-gradation obtained by giving degree $\bar{0}$ to the field $\mathbb{R}$ and degree $\bar{1}$ to the (co-) vectors of $\mathbb{R}^D.$}
generated by $\theta_\al,$ $1 \leq \al \leq D.$ This is the standard operation of tensoring a Lie algebra    ($\mf{g}$ say) by a graded associative one ($\Lambda$ here) -- the superLie bracket is given by
$[g\otimes \lambda,g' \otimes \lambda' ]_{super}$:=$[g,g'] \otimes \lambda \lambda',$ $g,g' \in \mf{g},
\lambda, \lambda' \in \Lambda .$
The natural $\mathbb{Z}$-gradation of $\Lambda\! \pa{\mathbb{R}^{D}}$ extends to this algebra by giving degree $0$ to all $\mathcal{P}E_{11}\pa{D}$ generators and will lead exactly to the V-degree.
Below we shall only distinguish the Lie bracket from the superbracket when the ambiguity will become annoying.

Note that in this construction, it is essential to restrict one's attention to the parabolic subalgebra $\mathcal{P}E_{11}\pa{D}$ in which the lowering generators $f_i$ associated with the starred nodes have been dropped.  Indeed, these $f_i$'s are not scalars, but transform in the representation dual to that of the corresponding $e_i$'s. 
To include all the $f_i$'s would necessitate introducing the dual exterior algebra generated by the $dx^\al$ dual to the $\theta_\al$ and this would change the construction.  It would be of interest to explore how far one can go in that direction.  We shall not do it here, defining in the end of the analysis the missing lowering generators of the resulting Borcherds superalgebras by assuming the existence of a Chevalley involution.

There are several different $\mf{gl}_D$ actions on the algebra $\mathcal{A} \equiv \mathcal{P}E_{11}\pa{D} \otimes \Lambda  \! \pa{\mathbb{R}^{D}}.$  First one can use the adjoint action of $\mf{gl}_D \subset \mathcal{A}$ on $\mathcal{A}.$  The $\theta_\alpha$'s are clearly inert under it since they do not contribute to the brackets.  The second action, that we shall call the natural action, coincides with the adjoint action on $\mathcal{P}E_{11}\pa{D}$ but transforms also the $\theta_\alpha$'s as ``vectors" (with the same abuse of terminology as in subsection 1.1 for ``forms").

\section{Subsuperalgebra of Invariants}
\label{pformalgebra}
\setcounter{equation}{0}

\subsection{Invariant generators}

We consider from now on the natural action of $\mf{gl}_D$ described in the previous paragraph.
Our  superalgebra  
 $\mathcal{A} := \mathcal{P}E_{11}\pa{D} \otimes \Lambda \! \pa{\mathbb{R}^{D}}$ provides a representation of $\mf{gl}_D$ which is completely reducible.  We denote by $\mathcal{A}_0$ the subspace containing the invariant elements in $\mathcal{A},$ i.e., the subspace of the trivial representations for the natural action of $\mf{gl}_D.$ It is a subsuperalgebra (subalgebra for short),  which contains the elements of $E_{11}$ properly saturated with $\theta_\alpha,$ or with indices properly contracted.  The central claim of this paper is that, for each spacetime dimension $D,$ $\mathcal{A}_0$ is (a truncated version of the parabolic subalgebra of) the Borcherds superalgebra 
 $\mf{V}_D$ considered in \cite{HenryLabordere:2002dk}. The truncation follows at the end of our computation from the finite dimensionality of the Grassmann algebra of parameters.

The elements in the invariant subalgebra $\mathcal{A}_0$ are the scalars, obtained by saturating completely upper and lower indices, and the forms contracted with the $\theta$'s, e.g., $E_1:=\frac{1}{3!} E^{\al \be \gamma} \theta_\al \theta_\be \theta_\gamma$ in eleven dimensions.  So, only the completely antisymmetric tensors survive.  Tensors with mixed Young symmetry are eliminated when saturating their indices with products of $\theta$'s. 

It is of interest to stress that $\mathcal{A}_0$ is a subalgebra of the superalgebra $\mathcal{P}E_{11}\pa{D} \otimes \Lambda \! \pa{\mathbb{R}^{D}}$ while the set of fully antisymmetric elements of $\mathcal{P}E_{11}\pa{D}$ does not define a subalgebra of $\mathcal{P}E_{11}\pa{D}.$  Indeed the bracket of two fully antisymmetric generators is not necessarily fully antisymmetric and so one needs to multiply by the $\theta$'s to get rid of pieces with mixed Young symmetry.  

There is, however, an alternative way of describing the set of fully antisymmetric elements of $\mathcal{P}E_{11}\pa{D}$ within $\mathcal{P}E_{11}\pa{D}$ without introducing superalgebras.  The elements of $\mathcal{P}E_{11}\pa{D}$ with at least two columns form an ideal $\mathfrak{I},$ this follows from the rules for computing tensor products of representations. The set of fully antisymmetric elements of $\mathcal{P}E_{11}\pa{D}$ can be identified with the quotient algebra $\mathcal{P}E_{11}\pa{D}/ \! \mathfrak{I}.$  This line of reasoning holds for   $\mf{gl}_D$ tensors but not  for $\mf{sl}_D$ ones and only if one considers density weights of the same sign, see for instance \cite{FH} .

Multiplying by the $\theta$'s automatically takes the quotient, in fact  it does also truncate the superalgebra to  degree $D.$ We are going to ignore this until the end and the application to Physics as otherwise the beauty of parabolic Borcherds superalgebras would be hidden. 

A possibility to avoid this complication could be to go beyond  $E_{11}$ to  $E_{n},$ $n \geq 12,$
or even to the infinite rank situation. Indeed the branching rules of the decomposition described by Fig. {\bf 1} above seem to stabilize as for  the already noticed similarity between   $E_{10}$ and  $E_{11}$ prescriptions. By so doing one would obtain exactly the full parabolic subalgebras of superBorcherds. What happens is that the $\mf{gl}_{D+k}$ invariants of $\mathcal{P}E_{11+k}\pa{D+k} \otimes \Lambda  \! \pa{\mathbb{R}^{D+k}}$ form an algebra that does not depend on k for large k. More information on this point is provided in Appendix B.

It is an exciting exercise to look for similar constructions keeping up to two or $p$ columns instead of just one, this might lead to a symmetry superalgebra "inbetween" the too large $E_{11}$ and the too small V-duality. 

In order to establish the assertion that $\mathcal{A}_0$ is a parabolic subalgebra of the Borcherds superalgebra $\mf{V}_D,$ we first construct the invariants associated with the simple root-vectors $e_i,$ the conjugate $f_i$'s  that are in $\mathcal{P}E_{11}\pa{D}$ and the Cartan generators $h_i.$ In the Cartan subalgebra of $E_{11},$ the invariant elements are the scalar generators of the Cartan subalgebra of the internal symmetry, as well as the trace $K_D,$ which must be traded for $ K_{11}$ in the case $D=2,$ since then as we saw $K_2$ is not independent from the Cartan generators of the internal U-duality algebra. 
When there are two starred nodes, an additional invariant exists, which may be taken to be $h_* = K^{11}_{\; \; 11}$ (for $D\! = \! 10_A$), or for instance $h'_* = K^{10}_{\; \; 10}$ (for D=9).

Out of the simple raising operators of $E_{11},$ one can construct invariants by saturating the form indices with $\theta_\alpha$'s (with no $\theta_\alpha$ needed for scalar generators).  From the list of covariant generators given in Subsection {\bf \ref{covariantizationZZZ}}, one thus gets Table {\bf \ref{InvGen}} of invariant generators. {}For the sake of conciseness, we omit the lowering generators $f_a$ of the internal U-duality , i.e., we consider only the ``Borel" (see below) subalgebra $\mathcal{B}_0$ of $\mathcal{A}_0.$  The lowering generators $F_a=f_a$  can be reintroduced using a Chevalley involution.

\subsection{Structure of invariant subalgebra}

We now study in more detail the structure of the invariant subalgebra $\mathcal{B}_0.$  We want to verify two properties: \begin{itemize}
\item $\mathcal{B}_0$ is generated by the invariant generators $\bar{H_A}$ (Cartan) and $E_A$ (positive simple root generators)  of Table {\bf \ref{InvGen}} with $A=1,...,12 \! - \! D.$\item The only relations on the generators of $\mathcal{B}_0$ are those that follow by taking traces or contracting with the $\theta$'s the covariant relations given in Subsection {\bf \ref{covariantizationZZZ}}.  There are no other relations if the truncation below degree $D$ is temporarily put aside.

\end{itemize}

\begin{table}
{
\begin{center}
\begin{tabular}{|c||c|c|}
\hline
\multicolumn{3}{|c|}{Invariant generators} \\
\hline \hline
& Bosonic& Fermionic \\
\hline
$D=11$ & $\bar{H_1} := K_{11} $ & $E_1 := \frac{1}{3!} E^{\alpha \beta \gamma} \theta_\alpha \theta_\beta \theta_\gamma$ \\ \hline
$D=10_A$ & $\bar{H}_1 := K_{11},$ $\bar{H}_2 \equiv K^{11}_{\; \; 11},$ & $E_1 := K^\alpha_{\; 11} \theta_\alpha$ \\ & $E_2 := \frac{1}{2!} E^{\beta \gamma \ts 11} \theta_\beta \theta_\gamma$ & \\
\hline
$D=10_B$ & $\bar{H}_1 := h_{10'},$ $\bar{H}_2 := K^\alpha_{\; \alpha},$ &  \\ & $E_1 := e = e_{10'},$ $E_2 := \frac{1}{2!} E^{\alpha \beta } \theta_\alpha \theta_\beta $ & \\ \hline
$D=9$ & $\bar{H}_1 := K_{11},$ $\bar{H}_2 := h_{10},$ $\bar{H}_3 :=K^{10}_{\; \; 10},$ &$E_1 := K^\alpha_{\; 10} \theta_\alpha,$  \\ & $E_2 := e_{10}$  & $E_3 :=  E^{\alpha \ts 10 \ts 11} \theta_\alpha $ \\
\hline
$1 \leq D \leq 8$ & $\bar{H}_1 := K_{11},$ $\bar{H}_n := h_{D+n-1},$ &$E_1 := K^\alpha_{\; D+1} \theta_\alpha$ \\ & $E_n := e_{D+n-1}$ & \\
\hline
\end{tabular}
\end{center}
}
\caption{The generators of the invariant subalgebra (in the last two lines, $n$ runs from $2$ to $12\! - \! D).$}
\label{InvGen}
\end{table}

We shall provide details in \cite{HJLFuture} and  give here informal arguments. 

\subsubsection{Generators of $\mathcal{B}_0$}
Let $\omega$ be an exterior form in $\mathcal{B}_0.$  Without loss of generality, we can consider the homogeneous case of degree $p.$ The form $\omega$ can be written as a linear combination of terms of the form $a^{\lambda_1 \cdots \lambda_p} \, \theta_{\lambda_1} \wedge \cdots \wedge \theta_{\lambda_p},$ where $a^{\lambda_1 \cdots \lambda_p}$ is a multicommutator of the covariant generators listed in Table {\bf \ref{Table1}} (if $p{>}0,$ we can assume that there is no $K^\al_{\; \be}$ in the commutator as these can be gotten rid of using the commutation relations of Subsection {\bf \ref{covariantizationZZZ}}).  Because the $\theta$'s anticommute, we can group them so as to make the forms of Table {\bf \ref{InvGen}} appear in the multicommutator, e.g., for D=11 the term $[E^{\lambda_1 \lambda_2 \lambda_3}\theta_{\lambda_2} \theta_{\lambda_3} \theta_{\lambda_4}, E^{\lambda_4 \lambda_5 \lambda_6}\theta_{\lambda_1}\theta_{\lambda_5} \theta_{\lambda_6}]_{super}$ is in $\mathcal{B}_0;$ it is equal to  $-[E^{\lambda_1 \lambda_2 \lambda_3}, E^{\lambda_4 \lambda_5 \lambda_6}] \, \theta_{\lambda_1} \theta_{\lambda_2} \theta_{\lambda_3} \theta_{\lambda_4} \theta_{\lambda_5} \theta_{\lambda_6}$ and thus is expressible in terms of the invariant generators as 
$-36\ts [E_1,E_1]_{super}.$

This shows that $\mathcal{B}_0$ is generated by the invariant generators of Table {\bf \ref{InvGen}}.  The superalgebra $\mathcal{A}_0$ is generated by these invariant generators together with the lowering generators $f_a$ of the $U$-duality group.

\subsubsection{Relations on the generators of $\mathcal{B}_0$}

\begin{table}
\begin{center}
{\footnotesize
\begin{tabular}{|c||c|c|}
\hline
\multicolumn{3}{|c|}{Relations} \\
\hline \hline
& Chevalley relations& Serre relations \\
\hline
$D=11$ & $[\bar{H}_1, E_1] = 3E_1 .$ & None \\ \hline
$D=10_A$ & \begin{tabular}{cc} $[\bar{H}_1, E_1] = 0,$ & $[\bar{H}_1, E_2] = 3 E_2,$ \\ $[\bar{H}_2, E_1] = -E_1,$ & $[\bar{H}_2, E_2] = E_2.$ \end{tabular}& $[E_1, E_1] = 0.$ \\
\hline
$D=10_B$ & \begin{tabular}{cc}$[\bar{H}_1, E_1] = 2 E_1,$ & $[\bar{H}_1, E_2] = - E_2,$ \\ $[\bar{H}_2, E_1] = 0,$ & $[\bar{H}_2, E_2] = 2E_2.$ \end{tabular} & $[E_1, [E_1, E_2]] = 0.$\\
\hline
$D=9$ & $ \bar{C} = \left(\begin{array}{ccc} 0 & 0 & 3\\ -1 & 2 & 0 \\ -1 & 1 & 1 \end{array} \right) \!  \!  .$ & \begin{tabular}{cc} $[E_1,E_1] = 0,$ & $[E_3, E_3] = 0,$ \\  $[E_2, E_3] = 0,$ &$[E_2, [E_2, E_1]] = 0.$ \end{tabular}\\
\hline
$D=8$ & $ \bar{C} = \left(\begin{array}{cccc} 0 & 0 & 0 & 3\\ -1 & 2 & -1 & 0 \\ 0 & -1 & 2 & 0 \\ -1 & 0 & 0& 2 \end{array} \right) \! \! .$  &\begin{tabular}{cc} $[E_1,E_1] = 0,$ & $[E_2, [E_2, E_1]] = 0,$ \\  $[E_3, E_1] = 0,$ &$[E_4, [E_4, E_1]] = 0,$ \\  $[E_2, [E_2, E_3]] = 0,$ &$[E_3, [E_3, E_2]] = 0,$ \\  $[E_2, E_4] = 0,$ &$[E_3, E_4] = 0.$ \end{tabular}\\
\hline
$1 \leq D \leq 7$ & \begin{tabular}{c} $\bar{C} = \left(\begin{array}{cc} 0 & u \\ v & a \end{array} \right) \! \! ,$ where \\ $u = \left(\begin{array}{cccc} 0 & 0 & \cdots & 3 \end{array}\right) \!  \!,$ \\ $v^t = \left(\begin{array}{cccc} -1 & 0 & \cdots & 0 \end{array}\right) \! \! ,$ \\ $a$ = Cartan matrix of $E_{11-D}.$\end{tabular}  &\begin{tabular}{cc} $[E_1,E_1] = 0$ (for $D{>}1$), & $[E_2, [E_2, E_1]] = 0,$ \\ \;  \; $[E_1, E_A] = 0 \; \; (A>2).$ & \end{tabular}\\
\hline
\end{tabular}
}
\caption{     Supercommutation relations among basic invariants.  \hfill  \mbox{   } The relation $[E_1,E_1]=0$ is not a compulsory Serre relation in $D=1$ since it is trivially satisfied by truncation. We denote the Chevalley relations in this new basis by $[\bar{H}_A, E_B] = \bar{C}_{AB} E_B.$
Similarly the Serre relations are encoded using a symmetric matrix $S$ by $\hbox{ad}_{E_A}^{\,\,\,(1 - \frac{2 S_{AB}}{S_{AA}})} (E_B) = 0 $ if  $S_{AA} >0$ and $A \not=B,$ \hfill 
as well as the relation $[E_A,E_B] = 0 $ if $S_{AB} = 0.$}
\label{relationsBorcherds}
\end{center}
\end{table}

Let us now turn to the relations among the invariant generators of $\mathcal{B}_0.$ Among these we have first the relations involving the Cartan generators which are easy to derive and which are collected in Table {\bf \ref{relationsBorcherds}} in the column ``Chevalley relations". Consider next an arbitrary relation $\mathcal{R}$ among the generators of $\mathcal{B}_0.$  Without loss of generality, we can assume that it contains none of the Cartan generators since these can be eliminated using repeatedly the Chevalley relations which we just discussed. The relation $\mathcal{R}$ is of course a consequence of the $E_{11}$ Chevalley-Serre relations and hence of their covariantized version but also of the finite dimensionality of the Grassmann algebra. 

Let us temporarily ignore the consequences of the finiteness of $D$ -- which kills every element of V-degree ${>}D$ -- and implement them at the end to make contact with Physics. It amounts to the fact that ``forms" can be of degree at most $D,$ but let us still implement antisymmetrization on pairs of indices of unspecified range.  The Christoffel $\epsilon$ totally antisymmetric tensor is not an invariant of  $\mf{gl}_D,$ only an invariant of $\mf{sl}_D.$ Insisting on full $\mf{gl}_2$ invariance restricts us to generators with  $E_9$ central charge equal to the $\mf{sl}_2$ tensorial rank.

So we shall consider at first only those relations that can be written as linear combinations of multicommutators, each of which involves at least one of the covariantized Serre relations of $\mathcal{P}E_{11}\pa{D}.$  There is no free index since each such relation is invariant (it expresses the vanishing of an element in the algebra generated by the invariant generators) and so all indices are saturated with the $\theta$'s.  Using the anticommutativity of the $\theta$'s, we can contract the ``fundamental form generators" with the $\theta$'s carrying the same indices.  In particular, this produces ``invariant Serre relations" covariantly saturated with the $\theta$'s, that is, relations involving the invariant generators only.

This follows from covariance under the full linear group and the absence of any possible upper index pair contraction. A covariant relation in $\mathcal{P}E_{11}\pa{D}$ involving  several fundamental forms must by linear covariance amount to  the vanishing of some Young projector. Once fully contracted with $\theta$'s it must be the same as the relation in $\mathcal{B}_0$ or $\mathcal{A}_0$ .
  
In other words, the relations in $\mathcal{B}_0$ beyond the Chevalley relations and the maximal degree truncation are those that follow from the covariantized Chevalley-Serre relations in $\mathcal{P}E_{11}\pa{D}$ by saturating the components of the fundamental forms with $\theta$'s carrying the same indices. These relations are collected in the column ``Serre relations" of Table {\bf \ref{relationsBorcherds}}.

Anticipating the identification with the Borcherds superalgebra, we have separated the relations into two groups:  those that will become the Chevalley relations and those that will become the Serre relations.

Note that some of the covariantized relations become identities when saturated with the $\theta$'s.  For example, in the case $D=11,$ the relation (\ref{commEE}), $\left[E^{\al \be \ga},E^{\lambda \mu \nu} \right] = \left[E^{[\al \be \ga},E^{\lambda \mu \nu]} \right],$ yields $[E_1,E_1]_{super} = [E_1,E_1]_{super}$ and is thus empty.  

There is in that case no "Serre" relation whatsoever on the fermionic generator $E_1.$ This might seem surprising at first sight as it is known that there is no $9$-form.  But the $9$-form is indeed absent, thanks to the Jacobi identity for the (graded) commutator, which reads $[[E_1, E_1]_{super}, E_1]_{super} = 0.$

The conclusion is that the superalgebra $\mathcal{B}_0$ is the superalgebra generated by the invariant generators of Table {\bf \ref{InvGen}} subject to the conditions collected in Table {\bf \ref{relationsBorcherds}}.  The superalgebra $\mathcal{A}_0$ has the additional lowering generators $f_a$ of the U-duality subalgebra and the corresponding Chevalley-Serre relations involving $f_a.$

\section{Identification of Borcherds Algebras}
\label{idBorcherds}
\setcounter{equation}{0}

To complete the analysis, we shall now show that the generators of the invariant algebra $\mathcal{A}_0$ just constructed and the relations among them define a parabolic subalgebra of a rank $12\!-\!D$ Borcherds superalgebra (truncated at form degree $> D$), which we identify through its Cartan matrix. In this section we shall ignore the V-degree truncation, related to the spacetime dimension -- it will be imposed at the end. For clarity we only treat again the Borel part $\mathcal{B}_0$ of $\mathcal{A}_0$ (and that of the whole Borcherds superalgebra), as the analysis can be straightforwardly extended to include the scalar internal lowering generators $f_a.$

The generators (Table {\bf \ref{Table1}}) of the invariant algebra obey the relations of Table {\bf \ref{relationsBorcherds}}.  The generators $\{E_A\}$ will be identified with the simple raising operators of the Borcherds superalgebra, while the $\{\bar{H}_A\}$'s will span the Cartan subalgebra $\mathfrak{H}$ of $\mathcal{B}_0$.  One has indeed in all cases $[\bar{H}_A, \bar{H}_B] = 0$ and $[H, E_A] = \alpha_A(H) E_A,$ $\forall H\in \mf{H},$ where the $\alpha_A$'s are linear forms on $\mathfrak{\mathfrak{H}}$ ($\alpha_A \in \mathfrak{H}^*$).  To identify this presentation as that of a Borcherds superalgebra we must show that all the relations among the $E_A$'s and the $\bar{H}_A$'s can be viewed as Chevalley-Serre relations (restricted to the Borel part), i.e., are captured by a Cartan matrix according to the rules defining Borcherds algebras.

Our strategy will follow three steps.  We shall first show that the relations involving the $E_A$'s can be identified as Serre relations of a Borcherds algebra provided one uses as Cartan matrix a symmetric matrix $S_{AB}$ whose explicit form depends on the dimension $D.$  This matrix is in fact not completely determined by the Serre relations alone.  We shall then show that one can change the basis of the abelian subalgebra $\mathfrak{H},$ from  $\{\bar{H}_A\}$ to $\{H_A\},$ in such a way that the commutators $[H_A, E_B]$ still obey Chevalley relations after replacing $\bar{C}_{AB}$  with a matrix $C_{AB}=S_{AB}.$  This will then prove our claim.

We also show that one can choose the remaining ambiguity in $S_{AB}=C_{AB}$ in such a way that $S_{AB} $ coincides with the Cartan matrix $A_{AB}$ considered for $D\geq 3$ in  \cite{HenryLabordere:2002dk}. 

Since the form of the Cartan matrix depends on $D,$ we proceed dimension by dimension. We shall consider explicitly the cases $D= 11$ and $D=10_A$ and then list the corresponding Cartan matrices for the other cases. Details will be available in \cite{HJLFuture}. We shall also treat the case $D=1$ explicitly at the end of the next section, because, while a (truncated) Borcherds algebra structure can be given, it is slightly different from the cases $2 \leq D \leq 7.$

\subsection{$D=11$}
In that case, there is only one $E_1,$ the $3$-form (which is fermionic), and one $\bar{H}_1,$ the trace $K_{11}.$ The graded commutator $[E_1,E_1]$ of the fermionic generator $E_1$ with itself is unconstrained. In order for the Serre relations to impose no relation on the graded commutator $[E_1,E_1],$ the one-by-one matrix $S$ cannot vanish but can otherwise be an arbitrary integer.  One has $[\bar{H}_1, E_1] = 3 \ts E_1 \not= 0,$ which agrees with what the Serre relations dictate.

By taking $H_1 = (-1/3) K_{11},$ one gets $[H_1,E_1] = -E_1,$ which takes the form of a Chevalley relation with the one-by-one matrix $C$ equal to $-1 \not= 0.$ This choice is made to recover the Cartan matrix of \cite{HenryLabordere:2002dk}. Any other choice is equivalent to it by mere rescaling of $H_1 .$ Thus, with this choice, the Borcherds superalgebra relevant to $D=11$ has Cartan matrix:
\begin{equation*} D=11: \; \; \; A = \big( -1 \big).  \end{equation*}

\begin{table}
\begin{center}
{\begin{tabular}{|c||c|}
\hline
\multicolumn{2}{|c|}{New Cartan generators of the $p$-form superalgebras ($D \geq 1$)} \\
\hline \hline
$D=11$ & $H_1=-\frac{1}{3}K_{11}.$ 
 \\ \hline
$D=10_A$ & $H_1=-\frac{1}{3}K_{11},$  $H_2=-\frac{1}{3}K_{11}+K^{11}_{\; \; 11}.$  \\
\hline
$D=10_B$ & $H_1=-\frac{1}{4}K'_{10}-\frac{1}{2}h_{11'},$  $H_2=h_{11'}$ ($h_{11'}=h_{10}$). \\
\hline
$D=9$ & $H_1=-\frac{2}{3}K_{11}+K^{11}_{\; \; 11},$  $H_2=-\frac{2}{3}K_{11}+K^{10}_{\; \; 10}+K^{11}_{\; \; 11}.$  \\
\hline
$D=8$ & $H_1=-K_{11}+K^{10}_{\; \; 10}+K^{11}_{\; \; 11}=-K_9,$  $H_i=h_{i+7}$ ($i=2,3,4$).  \\
\hline
$2 \leq D \leq 7$ & $H_1=-K_{D+1},$ $H_i=h_{i+D-1}$ ($2\leq i \leq 12-D$).  \\
\hline
$D=1$ & $H_A=h_A$ ($2\leq A \leq 11$), $H_1=-K^2_{\; \; 2}.$  \\
\hline
\end{tabular}
}
\caption{Cartan generators of the Borcherds $p$-form superalgebras for all $1\leq D \leq 11,$ after the change of basis in the Cartan subalgebras made to obtain symmetric Cartan matrices encoding both Serre and Chevalley relations ($A=C=S$). However other choices are possible for $H_1$ in $D=1.$}
\label{NewCartan}
\end{center}
\end{table}

\subsection{$D=10_A$}
In that case, there are two $E_i$'s (see Table {\bf \ref{InvGen}}) and two Cartan generators $\bar{H}_i.$  The relations among $E_1$ and $E_2$ are just $[E_1,E_1] = 0$ (see Table {\bf \ref{relationsBorcherds}}) and can be viewed as Serre relations provided the matrix $S_{ij}$ fulfills \begin{enumerate} \item $S_{11} = 0$ (in order to have $[E_1,E_1] = 0$), \item $S_{22}\leq 0,$ $S_{12} <0,$ $S_{21} <0$ (in order to avoid the relations $[E_1,E_2]=0,$  $[E_2,E_1]=0$). \end{enumerate}  The commutation relations of the $\bar{H}_A$'s with the $E_B$'s define a matrix $\bar{C}$ that is invertible.  By the linear redefinitions $H_1 = -\frac{1}{3} \bar{H}_1$ and $H_2 = -\frac{1}{3} \bar{H}_1 + \bar{H}_2,$ one gets a matrix $C_{AB}$ equal to \begin{equation} D=10_A: \; \; \; C = \left(\begin{array}{cc} 0 & -1 \\ -1 & 0 \end{array} \right)  \label{AofD=10A}\end{equation} and hence in the class of the matrices determined by the Serre relations.  This particular choice of $H_i$'s actually 
 yields the maximum possible values for $S_{22},$ $S_{12}$ and $S_{21}$ compatible with the Serre requirements.  The resulting matrix is symmetric and corresponds to the Cartan matrix found in \cite{HenryLabordere:2002dk}. Thus, the Borcherds superalgebra relevant to $D=10_A$ has Cartan matrix $A = C$ given by (\ref{AofD=10A}).

\subsection{Results for $D \geq 2$ -- Cartan matrices and Dynkin diagrams}

The same analysis applies to all spacetime dimensions $D \geq 2$. In each case, we find that the generators and relations can be cast in the Borcherds superalgebra form.  Furthermore the ambiguities in the $D\geq 3$ cases can be naturally resolved by a change of basis of the Cartan subalgebras in such a way that the resulting Cartan matrix coincides with the one of \cite{HenryLabordere:2002dk}. The Cartan generators of the new basis are listed in Table {\bf \ref{NewCartan}}. From $D=8$ down to $D=2$ we see that the first generator becomes $H_1=-K_{D+1}.$ 
\bigskip
The Cartan matrices and Dynkin diagrams that we obtain are collected in Table {\bf \ref{Cartan-Dynkin}}, where we follow the usual conventions used for example by \cite{HenryLabordere:2002dk} :
\begin{picture}(14,14)
\thicklines
\put(7,4){\circle{14}}
\end{picture}
means a bosonic root of length $2$ ($A_{AA}=2$), \begin{picture}(14,14)
\thicklines
\put(7,4){\circle{14}}
\put(2,-1){\line(1,1){10}}\put(2,9){\line(1,-1){10}}
\end{picture}
a bosonic root of length $0$ ($A_{AA}=0,$ $A$ bosonic),
\begin{picture}(14,14)
\thicklines
\put(7,4){\circle*{14}}
\end{picture}
a fermionic root of length $0$ ($A_{AA}=0,$ $i$ fermionic), 
\begin{picture}(14,14)
\thicklines
\put(7,4){\circle{14}}
\put(7,4){\circle*{10}}
\end{picture}
a fermionic root of length $\leq -1,$ and 
\begin{picture}(14,14)
\thicklines
\put(7,4){\circle{14}}
\put(7,4){\circle{13}}
\put(7,4){\circle{12}}
\put(7,4){\circle{11}}
\put(7,4){\circle{10}}
\end{picture}
a fermionic root of length $1,$ the number of lines between simple roots being the opposite of the off diagonal element of the symmetrized Cartan matrix.

\bigskip
Note that the imaginary roots with $A_{AA} = 0$ would have length one when measured through the intersection matrix between divisors on del Pezzo surfaces  \cite{HenryLabordere:2002dk}.

Note also that one easily goes from the Borel subalgebra explicitly exhibited here to the parabolic subalgebra containing also the $f_a$'s of the internal duality algebra $E_{11-D}.$  The commutation relations of the lowering generators $f_a$ with $H_A,$ $E_B$ and between themselves are manifestly compatible with the Cartan matrix.  One can then consider the full Borcherds algebra using the Chevalley involution to introduce the missing lowering generators. The V-gradation is defined by giving a degree to each root generator, as indicated in Table {\bf \ref{Cartan-Dynkin}}.

\begin{table}
\begin{center}
{\scriptsize
\begin{tabular}{|c||c|c|}
\hline
\multicolumn{3}{|c|}{Borcherds algebras for maximal supergravities ($D \geq 1$)} \\
\hline \hline
& Symmetric Cartan matrix A=C& \hspace{2.5cm} Dynkin diagram \hspace{2.5cm} \\
\hline
$D=11$ & $A = (-1) $ & \begin{picture}(10,20)
\thicklines
\put(0,7){\circle{14}}
\put(0,7){\circle*{10}}
\put(-15,7){\makebox(0,0){$3$}}
\end{picture} \\ \hline
$D=10_A$ & $A = \left(\begin{array}{cc} 0 & -1 \\ -1 & 0 \end{array} \right)$  & \begin{picture}(42,20)
\thicklines
\put(0,2){\circle*{14}}
\put(-15,2){\makebox(0,0){$1$}}
\put(42,2){\circle{14}}
\put(57,2){\makebox(0,0){$2$}}
\put(7,2){\line(1,0){28}}
\put(37,-3){\line(1,1){10}}
\put(37,7){\line(1,-1){10}}
\end{picture} \\
\hline
$D=10_B$ & $A = \left(\begin{array}{cc} 0 & -1 \\ -1 & 2 \end{array} \right)$ & \begin{picture}(42,20)
\thicklines
\multiput(0,2)(42,0){2}{\circle{14}}
\put(7,2){\line(1,0){28}}
\put(-15,2){\makebox(0,0){$2$}}
\put(-5,-3){\line(1,1){10}}
\put(-5,7){\line(1,-1){10}}
\end{picture}\\
\hline
$D=9$ & $ A = \left(\begin{array}{ccc} 0 & -1 & 0\\ -1 & 0 & -1 \\ 0 & -1 & 2 \end{array} \right) $ & \begin{picture}(52,50)
\thicklines
\put(84,-3){\circle{14}}
\put(42,-3){\circle*{14}}
\put(27,-3){\makebox(0,0){$1$}}
\put(0,39){\circle*{14}}
\put(-15,39){\makebox(0,0){$1$}}
\put(49,-3){\line(1,0){28}}
\put(40,2){\line(-1,1){34}}
\end{picture}\\
\hline
$D=8$ & $ A = \left(\begin{array}{cccc} 0 & -1 & 0 & -1\\ -1 & 2 & -1 & 0 \\ 0 & -1 & 2 & 0 \\ -1 & 0 & 0& 2 \end{array} \right) $  &\begin{picture}(84,60)
\thicklines
\multiput(42,-8)(42,0){2}{\circle{14}}
\put(0,-8){\circle*{14}}
\put(-15,-8){\makebox(0,0){$1$}}
\put(0,34){\circle{14}}
\put(7,-8){\line(1,0){28}}
\put(49,-8){\line(1,0){28}}
\put(0,-1){\line(0,1){28}}
\end{picture}\\
\hline
$2 \leq D \leq 7$ & $A = \left(\begin{array}{ccccc} 0 & -1 & 0 &\cdots & 0\\ -1 & 2 & a_{23} & \cdots & a_{2 \ts 12-D} \\ 0 & a_{32} & \cdots & \cdots & a_{3 \ts 12-D} \\ \vdots & \vdots & \vdots & \vdots & \vdots \\ 0 & a_{12-D \ts 2} & \cdots & \cdots & 2 \end{array} \right)$  &{\tiny \begin{picture}(140,60)
\thicklines
\put(0,-17){\circle*{14}}
\put(-15,-17){\makebox(0,0){$1$}}
\multiput(42,-17)(42,0){4}{\circle{14}}
\put(84,25){\circle{14}}
\put(7,-17){\line(1,0){28}}
\put(91,-17){\line(1,0){28}}
\put(49,-17){\dashbox{3}(28,0)}
\put(84,-10){\line(0,1){28}}
\put(133,-17){\line(1,0){28}}
\end{picture}}\\
\hline
$D=1$ & $A = \left(\begin{array}{ccccc} 1 & -1 & 0 &\cdots & 0\\ -1 & 2 & a_{23} & \cdots & a_{2 \ts 11} \\ 0 & a_{32} & \cdots & \cdots & a_{2 \ts 11} \\ \vdots & \vdots & \vdots & \vdots & \vdots \\ 0 & a_{11 \ts 2} & \cdots & \cdots & 2 \end{array} \right)$  &{\tiny \begin{picture}(140,60)
\thicklines
\put(0,-17){\circle{14}}
\put(0,-17){\circle{13}}
\put(0,-17){\circle{12}}
\put(0,-17){\circle{11}}
\put(0,-17){\circle{10}}
\put(-15,-17){\makebox(0,0){$1$}}
\multiput(42,-17)(42,0){4}{\circle{14}}
\put(84,25){\circle{14}}
\put(7,-17){\line(1,0){28}}
\put(91,-17){\line(1,0){28}}
\put(49,-17){\dashbox{3}(28,0)}
\put(84,-10){\line(0,1){28}}
\put(133,-17){\line(1,0){28}}
\end{picture}}\\
\hline
\end{tabular}
}
\caption{Cartan matrices and Dynkin diagrams for Borcherds algebras of maximal supergravities, with the choice of Cartan generators of table 6. The block matrix $(a_{ab})$  $a\neq 1 ,  b\neq 1,$ is the Cartan matrix of $E_{11-D}$ for $D\leq8$ or $D \! = \! 10_B.$ The number next to the node indicates the V-degree of the associated root generator if it's not $0.$}
\label{Cartan-Dynkin}
\end{center}
\end{table}

\section{New features and results for D=2 and D=1}
\label{LOW}
\setcounter{equation}{0}
\subsection{$D=2$}

As announced above, our process of tensoring by $\Lambda \! \pa{\mathbb{R}^D}$ and selecting invariants does also yield in $D=2$ a truncation of a well-defined Borcherds superalgebra. It has $E_9$ as its degree-$0$ subalgebra, and its basic representation in degree $1.$ Contrary to the $D \geq 3$ cases, the degree truncation remains infinite-dimensional.

The generalized  Cartan matrix and Dynkin diagram are those of $E_{10},$ but with a $0$ in place of a $2$ as entry $\pa{1,1}$ of the matrix, and with the corresponding first Dynkin node fermionic.

The central charge of $E_9$ equals the scaling generator $K_2$ in $E_{11}$.

\subsection{$D=1$}

At first glance the situation for $D=1$ is the same as for $2 \leq D \leq 7$ : we have $11$ raising operators, $E_1 \equiv K^\alpha_{\; D+1} \theta_\alpha=K^1_{\; 2} \theta_1$ (a $1$-form) and $E_A = e_{A}$ ($2 \leq A \leq 11$), and $11$ Cartan generators $\bar{H}_1= K_{11},$ $\bar{H}_A = h_{A}$ ($2 \leq A \leq 11$), with $\bar{H}_A$  and $E_A = e_{A}$ ($2 \leq A \leq 11$) being the fundamental raising and Cartan generators of the relevant "$U$-duality algebra" $\mathfrak{U}=E_{10}.$ 

The matrix $\bar{C}$ encoding the Chevalley relations for the $\bar{H}_A$ is $$ \bar{C} = \left( \begin{smallmatrix}
       0 & 0 & 0& 0 & 0 & 0 & 0 & 0 & 0 & 0 & 3
\\-1& 2& -1& 0 & 0 & 0 & 0 & 0 & 0 & 0 & 0
\\0 &-1& 2 & -1& 0 & 0 & 0 & 0 & 0 & 0 & 0
\\0 &0 & -1& 2 & -1& 0 & 0 & 0 & 0 & 0 & 0
\\0 &0 & 0 & -1& 2 & -1& 0 & 0 & 0 & 0 & 0
\\0 &0 & 0 & 0 &-1 & 2 & -1& 0 & 0 & 0 & 0
\\0 &0 & 0 & 0 & 0 & -1& 2 & -1& 0 & 0 & 0
\\0 &0 & 0 & 0 & 0 & 0 &-1 & 2 & -1& 0 & -1
\\0 &0 & 0 & 0 & 0 & 0 & 0 &-1 & 2 & -1& 0
\\0 &0 & 0 & 0 & 0 & 0 & 0 & 0 &-1 & 2 & 0
\\0 &0 & 0 & 0 & 0 & 0 & 0 & -1& 0 & 0 & 2
 \end{smallmatrix} \right) \! \! ,$$
but we must find a basis $\{H_A\}$ for which the Chevalley relations are encoded in a symmetric matrix of the type 
$$ S_{z}=\left( \begin{smallmatrix}
  z & -1 & 0& 0 & 0 & 0 & 0 & 0 & 0 & 0 & 0
\\-1& 2& -1& 0 & 0 & 0 & 0 & 0 & 0 & 0 & 0
\\0 &-1& 2 & -1& 0 & 0 & 0 & 0 & 0 & 0 & 0
\\0 &0 & -1& 2 & -1& 0 & 0 & 0 & 0 & 0 & 0
\\0 &0 & 0 & -1& 2 & -1& 0 & 0 & 0 & 0 & 0
\\0 &0 & 0 & 0 &-1 & 2 & -1& 0 & 0 & 0 & 0
\\0 &0 & 0 & 0 & 0 & -1& 2 & -1& 0 & 0 & 0
\\0 &0 & 0 & 0 & 0 & 0 &-1 & 2 & -1& 0 & -1
\\0 &0 & 0 & 0 & 0 & 0 & 0 &-1 & 2 & -1& 0
\\0 &0 & 0 & 0 & 0 & 0 & 0 & 0 &-1 & 2 & 0
\\0 &0 & 0 & 0 & 0 & 0 & 0 & -1& 0 & 0 & 2
 \end{smallmatrix} \right) \! \! .$$  Here $z$ can be any real number we want, as in this case $E_1$ has only one term $E_1=K^1_{\; 2} \theta_1,$ and so we don't need to impose the relation $[E_1,E_1]=0,$ it is automatically satisfied by the degree truncation. However, one finds that we can't take our usual type of Cartan matrix with a $0$ in the upper-lefthand corner : indeed, when $z=0$ the determinant of $S_{0}$ is zero (this comes from the fact\footnote{This choice of Cartan matrix would give as in higher dimensions $H_1=-K_{D+1}=-K_2,$ but $K_2$ belongs to $E_9,$ so this can't yield a basis of the Cartan subalgebra of $E_{11}$ together with the $h_a$'s, $2\leq a \leq 11.$} that the determinant of the Cartan matrix of $E_9$ is $0$) while the determinant of $\bar{C}$ isn't, and thus we can't find a basis yielding this matrix for its Chevalley relations. 

But we can nonetheless obtain any value $z \neq 0,$ as in that case the determinant of $S_{z}$ is no more zero. Hence there exists many bases $\{H_A\}$ of the Cartan subalgebra yielding $S_{z}$ as matrix of the Chevalley relations, namely 
$H_{1}\pa{z}=\pa{z-1}K^1_{\; 1} - K_{\; 2}^2$ and $H_A=\bar{H_A},$ $2 \leq A \leq 11.$ Here we chose, as in higher dimensions, a  symmetric matrix $S,$ it was not forced upon us. But we are left with an ambiguity in the choice of the first Cartan vector coming from the freedom on the choice of $z \neq 0.$ A natural choice is to take $z=1,$ so that $H_1= -K^2_{\; 2}$ and the Cartan matrix $A=S_{1}=C.$

Here, once truncated to respect the dimension $D=1,$ the $\mathbb{Z}$-gradation of our algebra has only two non trivial homogenous components, which are also the $\bar{0}$  (bosonic) and $\bar{1}$ (fermionic) components of the $\mathbb{Z}_2$-gradation. We can readily see that the Deg $0$ part of our Borcherds algebra, the scalars, is formed by the Borel of $E_{10}$ and that its Deg $1$ part, i.e., the $1$-forms, is formed by a highest weight representation of the hyperbolic algebra $E_{10}.$ Of course, as for $D=2,$ the dimension of the (truncated) algebra is infinite.

So, again for $D=1,$ by our process of tensoring and selecting the invariants, we have obtained  a truncation of a Borcherds superalgebra, we hope to return to this and investigate what could be a true symmetry. However, if we ignore the degree truncation and consider only the relations coming from the $\mf{gl}_1$-covariantisation of the relations of $E_{11},$ this does not define a Borcherds superalgebra, for the reasons explained above and linked to the affine character of $E_9.$ However, if we are only interested in the truncation, it can be seen as coming from the real Borcherds superalgebra of Table {\bf 7}.

\section{Conclusions}
\label{conclusions}
\setcounter{equation}{0}

In this paper, we have demonstrated that the reduction of $E_{11}$ to the $p$-form sector of maximal supergravities in $D \geq 3$ dimensions leads to the $V$-duality symmetries described by Borcherds superalgebras.  We have recovered, in particular, the Cartan matrices of \cite{HenryLabordere:2002dk} from the Cartan matrix of $E_{11},$ thereby explaining the harmony between the result of the Borcherds and the $E_{11}$ methods for calculating the $p$-form content. The coincidence with the constraints of supersymmetry remains a mystery. We have also proved that $E_{11}$ implies that the $D=2$ case is also encoded in a Borcherds superalgebra and that the analysis extends all the way down to $D=1.$

Even though we have clarified the connection between $E_{11}$ and the Borcherds superalgebras used earlier in \cite{HenryLabordere:2002dk} to describe economically the supergravity $p$-form spectra, many questions remain open. The precise role of $E_{11}$ remains mysterious in that many fields at higher levels are still waiting for a precise physical interpretation.  It appears to be quite magical, however, that among the higher level fields, one finds always, for any spacetime dimension, fields that precisely reduce to the fields necessary to describe deformations and top forms \cite{Riccioni:2007au,Bergshoeff:2007qi}. Anyway, as the Borcherds character of our symmetries allow us to use the denominator formula, we now have an easy -- at least for high dimensions -- tool to determine exactly the roots corresponding, not only to propagating forms, but also to de-forms and top forms.

It would be of interest to extend this work to theories with other internal duality groups like $D_8^{+++}$ or $B_8^{+++}.$  Similarly, one would like to understand better the important cases $D=2$ and $D=1$ and in particular their top forms. Another important problem is to understand the interplay of self-duality and the exchange symmetry about dimension $\frac{D-2}{2}$ on the one hand, and the ordinary Chevalley involution exchanging opposite Borel subalgebras on the other hand. It is hoped to return to these questions in the future \cite{Future}.

\section*{Acknowledgements} M. Henneaux thanks the Laboratoire de Physique Th\'eorique de l'Ecole Normale Sup\'erieure for kind hospitality  and gratefully acknowledges support from the Alexander von Humboldt Foundation through a Humboldt Research Award.  The work of M. Henneaux is partially supported by IISN - Belgium (conventions 4.4511.06 and 4.4514.08), by the Belgian Federal Science Policy Office through the Interuniversity Attraction Pole P6/11 , by the ERC Advanced Grant SyDuGraM and by the ``Communaut\'e Fran\c{c}aise de Belgique" through the ARC program. B. Julia would like to dedicate this work to the memory of his high school Mathematics teacher Denis Gerll who year after year communicated to his class his enormous enthusiasm. He is grateful to GGI in Firenze for stimulating hospitality and an enjoyable program and to SNS Pisa for interest and hospitality.  J. Levie wants to thank the Laboratoire de physique math\'ematique des interactions fondamentales de l'Universit\'e  libre de Bruxelles for its courteous hospitality.

\appendix

\section{Borcherds or BKM superalgebras}
\setcounter{equation}{0}
\label{DefinitionBorcherds}

In this appendix, we provide some basic notions on Borcherds superalgebras.  For more information, see \cite{bor,Kac}.
\vspace{.2cm}

\noindent
{\bf Definition:} Let $I = \{1, \cdots, N \}$ be an index set with both ``bosonic" and ``fermionic" indices.  Let $S \subset I$ be the subset of fermionic indices.  A generalized symmetric Cartan matrix $C = (a_{ij})$ ($i \in I$) of a supersymmetric Borcherds (``generalized Kac-Moody") algebra
is a non-degenerate symmetric matrix ($a_{ij} = a_{ji}$) with the following properties:
\begin{itemize}
\item $a_{ii}$ can be $<0,$ $0$ or $>0;$
\item $a_{ij} \leq 0$ if $i \not=j;$
\item If $a_{ii} >0,$ then $\frac{2 a_{ij}}{a_{ii}} \in \mathbb{Z}$ for all $j \in I;$
\item More stringently if $a_{ii} >0$ and $i \in S,$ then $\frac{ a_{ij}}{a_{ii}} \in \mathbb{Z}$ for all $j \in I.$
\end{itemize}

The Borcherds superalgebra $\mathcal{A}$ associated with the generalized Cartan matrix $a_{ij}$ is generated by $3N$ generators $\{h_i, e_i, f_i\}$ ($i = 1, \cdots, N$) subject to the following relations
\begin{eqnarray} && [h_i,h_j] = 0, \label{HH2} \\ &&[h_i, e_j] = a_{ij} e_j, \; \; [h_i, f_j] = -a_{ij} f_j , \; \; [e_i, f_j] = \delta_{ij} h_i, \label{Chev2}\\
&& \hbox{deg }e_i = 0 = \hbox{deg }f_i \; \; \hbox{if $i\notin S$}\, , \; \; \; \hbox{deg }e_i = 1 = \hbox{deg }f_i   \; \;  \hbox{if $i\in S,$} \\
&& \left(\hbox{ad}_{e_i}\right)^{1 - \frac{2 a_{ij}}{a_{ii}}}e_j = \left(\hbox{ad}_{f_i}\right)^{1 - \frac{2 a_{ij}}{a_{ii}}}f_j = 0  \; \; \hbox{if $a_{ii} >0$ and $i \not=j,$} \label{Serre21} \\
&& \hbox{furthermore }  [e_i,e_j] = 0 = [f_i, f_j] \; \; \; \hbox{ if $a_{ij} = 0$} \label{Serre32} .
\end{eqnarray}
Relations (\ref{HH2}) and (\ref{Chev2}) are the Chevalley relations, relations (\ref{Serre21}) and (\ref{Serre32}) are the Serre relations\footnote{In \cite{Ray}, the condition $\left(\hbox{ad}_{e_i}\right)^{1 - \frac{ a_{ij}}{a_{ii}}}e_j = 0 , \; \; \left(\hbox{ad}_{f_i}\right)^{1 - \frac{ a_{ij}}{a_{ii}}}f_j = 0 $ when
$i \in S,$ $a_{ii}>0$ and $i \not=j$ is imposed. As the left-hand sides of these relations do not define ideals that intersect trivially the Cartan subalgebra, this appears to be incorrect.  The condition that $\frac{ a_{ij}}{a_{ii}}$ should be in $\mathbb{Z}$ when $i$ is a fermionic index such that $a_{ii}>0$ is essential, however.}.

So, the idea behind the extension to Borcherds superalgebras (with respect to standard Kac-Moody superalgebras) is that one relaxes some of the conditions on the matrix $a_{ij},$ which is now allowed to have diagonal elements which are $\leq 0.$  When $a_{ii} \leq 0,$ the corresponding simple root is imaginary (contrary to a Kac-Moody algebra where all simple roots are real) and there is no restriction on $\frac{a_{ij}}{a_{ii}}.$

When $a_{ii} >0,$ the (negative) integers $\frac{2 a_{ij}}{a_{ii}}$ appearing in the Serre relations are called the Cartan integers.  The Cartan integer $\frac{2 a_{ij}}{a_{ii}}$ is even when $i$ is fermionic.

For a Borcherds superalgebra, the triangular decomposition still holds and roots can be defined in the same manner as for Kac-Moody algebras.  However, there are new features.  It is still true that if $a_{ii}>0,$ the following properties hold
\begin{itemize}
\item If $i$ is a bosonic index, then the Lie subalgebra $$ S_i = \mathbb{C} f_i \oplus \mathbb{C} h_i \oplus \mathbb{C} e_i $$ is isomorphic to $\mathfrak{sl}_2$;
\item If $i$ is a fermionic index, then the Lie sub-superalgebra $$ S_i = \mathbb{C} [f_i, f_i] \oplus \mathbb{C} f_i \oplus \mathbb{C} h_i \oplus \mathbb{C} e_i \oplus \mathbb{C} [e_i, e_i] $$ is isomorphic to $\mathfrak{osp} (1\vert 2)$;
\item The Kac-Moody superalgebra decomposes into finite-dimensional representations of $S_i$ (because of the Serre relations).
\end{itemize}
But these properties no longer hold when $a_{ii} \leq 0$:
\begin{itemize}
\item If $a_{ii} = 0,$ the sub(super)algebra $$\mathbb{C} f_i \oplus \mathbb{C} h_i \oplus \mathbb{C} e_i$$ is isomorphic to the Heisenberg (super)algebra;
\item If $a_{ii} <0,$ the subalgebra $$ S_i = \mathbb{C} f_i \oplus \mathbb{C} h_i \oplus \mathbb{C} e_i $$ and the sub-superalgebra $$ S_i = \mathbb{C} [f_i, f_i] \oplus \mathbb{C} f_i \oplus \mathbb{C} h_i \oplus \mathbb{C} e_i \oplus \mathbb{C} [e_i, e_i] $$ are still isomorphic to $\mathfrak{sl}_2$ and $\mathfrak{osp} (1 \vert 2),$ respectively, but the Borcherds superalgebra $\mathcal{A}$ contains infinite-dimensional representations of $S_i.$
\end{itemize}

\section{Lifting temporarily the restriction to form degrees less than $D$}
\setcounter{equation}{0}

In this paper, we are interested in the algebra of $p$-forms in any spacetime dimension $D \leq 11 .$  We have shown that this algebra is generated by the generators of table 4 subject to the conditions of table 5 {\it and} the extra condition that the form degree $p$ is bounded by $D,$ $p \leq D$.

Technically, this extra condition arises because there are $D$ anticommuting $\theta_\alpha$'s and so any expression involving more than $D$ $\theta$'s identically vanishes.

Our claim is that this algebra is the restriction to form degree $\leq D$ of the parabolic subalgebra of the Borcherds algebra $\mathfrak{V}_D$ derived in the text.  In other words, if one drops the form degree restriction, the $p$-form algebra and this parabolic subalgebra coincide. 

Suspending temporarily  the form-degree restriction necessary to reach  $\mathfrak{V}_D$ might appear artificial.  We show in this appendix that this can naturally be viewed as replacing $E_{11}$ by $E_{11+k}$ and taking the limit of large $k.$

To see this, consider first the subsuperalgebra of $ \mathfrak{gl}_{D+1}$-invariants in dimension $D+1$ for $E_{12}.$  We call ``$0$" the additional node of $E_{12}$  keeping the same labels for the other nodes of the Dynkin diagram.  So, in particular, the exceptional node is still called ``$11$".   

Comparing the reduction to $D+1$ dimensions for $E_{12}$ to the reduction to $D$ dimensions for $E_{11}$ amounts to assuming the same number $11 \! - \! D= 12\!  - \! D \! - \! 1$ of internal dimensions. Our crucial point  follows immediately from that observation, which implies that the generators of the subsuperalgebra of $ \mathfrak{gl}_{D+1}$ invariants  for $E_{12}$ are {\it formally exactly the same} as the $ \mathfrak{gl}_{D}$ invariant ones for $E_{11}.$  The only difference is that the index $\alpha$ in $E_1 = K^\alpha_{\; D+1} \theta_\alpha$ now runs from $0$ to $D,$ i.e., can take the additional value $0.$  Similarly, the relations among the invariant generators listed in table 5 are unchanged.  Hence, the only difference between the $p$-form algebra associated with $E_{12}$ in $D+1$ dimensions and that associated with $E_{11}$ in $D$ dimensions is that the form-degree truncation now occurs at degree $D+1.$

Similarly, if one were to consider the reduction of $E_{13}$ to $D+2$ dimensions (calling the additional $E_{13}$-node ``$-1$"), one would get exactly the same collection of $0$-forms, $1$-forms, $2$-forms, ... up to degree $D$ (comparing with $E_{11}$) or $D+1$ (comparing with $E_{12}$), but now there would also be $(D+2)$-forms.   Furthermore, for any value of $p,$  the $p$-forms appearing in $E_{11}$, $E_{12}$ or $E_{11+k}$ ($p \leq D$), being formally equal, do transform in the same representation of the U-duality internal algebra (which does not depend on $k$ since we keep the number of internal dimensions constant) .

The pattern is now obvious: to temporarily hold the truncation to form degree $D$, one simply goes to $E_{11+k}$ and considers $k$ arbitrarily large, i.e., the infinite rank situation.  This yields the form algebra generated by the generators of table 4 subject {\it only} to the conditions in table 5, with no form-degree restriction.  It is this ``universal" $p$-form algebra that is identical with the parabolic subsuperalgebra of the Borcherds superalgebra described in the text. The physical $p$-form algebra is obtained by making the form-degree truncation $p \leq D,$ which we have chosen to postpone till  the end in order to reveal the underlying Borcherds structure.

\end{document}